%% file: PaperForReview.tex
\documentclass[10pt,twocolumn,letterpaper]{article}
\pdfoutput=1
\usepackage{cvpr}              % To produce the CAMERA-READY version

\makeatletter
\@namedef{ver@everyshi.sty}{}
\makeatother

\usepackage{graphicx}
\usepackage{amsmath}
\usepackage{amssymb}
\usepackage{booktabs}
\usepackage{dsfont}

\usepackage{lipsum}

\usepackage[pagebackref,breaklinks,colorlinks]{hyperref}

\usepackage[capitalize]{cleveref}
\crefname{section}{Sec.}{Secs.}
\Crefname{section}{Section}{Sections}
\Crefname{table}{Table}{Tables}
\crefname{table}{Tab.}{Tabs.}

\usepackage[table,xcdraw]{xcolor}
\usepackage{amsmath}
\usepackage{epsfig}
\usepackage{graphics}
\usepackage{multirow}
\usepackage{subcaption}
\usepackage{mwe}
\usepackage{makecell}
\usepackage{url}
\usepackage{soul}
\usepackage[inkscapelatex=true]{svg}
\newcommand{\comment}[1]{}

\usepackage[accsupp]{axessibility} % Improves PDF readability for those with disabilities.

\usepackage[normalem]{ulem} %strikethrough

\usepackage{xspace}

\usepackage{hyperref} %Used for \autoref
\usepackage{bm} %boldmath
\usepackage{amsmath} %text in equation
\usepackage{amssymb}
\usepackage{cleveref}

\usepackage{adjustbox} %table stuff
\usepackage{multirow}
\usepackage[para,online,flushleft]{threeparttable}
\usepackage{booktabs}
\usepackage[]{tikz}
\usepackage{algorithmicx}
\usepackage{algorithm}
\usepackage[noend]{algpseudocode}

\usepackage{color}

\newcommand{\figref}[1]{ Fig. \ref{#1}}
\newcommand{\tabref}[1]{ Table \ref{#1}}

\let\oldtextbf\textbf
\renewcommand{\textbf}[1]{\oldtextbf{\boldmath #1}}

\usepackage{graphicx}
\usepackage{tikz}

\usepackage{float}

\def\paperID{23} % *** Enter the Paper ID here % *** Enter the EarthVision Paper ID here (taken from CMT)
\def\confName{EarthVision}
\def\confYear{2024}

\begin{document}

%%%%%%%%% TITLE - PLEASE UPDATE
\title{Implicit Assimilation of Sparse In Situ Data \\ for Dense \& Global Storm Surge Forecasting}

\author{Patrick Ebel$^{*}$\\
{\tt\small patrick.ebel@esa.int}
% For a paper whose authors are all at the same institution,
% omit the following lines up until the closing ``}''.
% Additional authors and addresses can be added with ``\and'',
% just like the second author.
% To save space, use either the email address or home page, not both
\and
Brandon Victor$^{\dag}$\\
{\tt\small b.victor@latrobe.edu.au}
\and
Peter Naylor$^{*}$\\
{\tt\small peter.naylor@esa.int}
\and
Gabriele Meoni$^{*, \circ}$\\
{\tt\small gabriele.meoni@esa.int}
\and
Federico Serva$^{\ddag}$\\
{\tt\small federico.serva@terrarum.eu}
\and
Rochelle Schneider$^{*}$\\
{\tt\small rochelle.schneider@esa.int}
\and
{\small $^{*}$European Space Agency, $\Phi$-lab  \quad $^{\dag}$ La Trobe University \quad  $^{\circ}$European Space Agency, ACT \quad $^{\ddag}$ Consiglio Nazionale delle Ricerche
}
}

\maketitle

% As a general rule, do not put math, special symbols or citations
% in the abstract or keywords.

\begin{abstract}
Hurricanes and coastal floods are among the most disastrous natural hazards. Both are intimately related to storm surges, as their causes and effects, respectively. However, the short-term forecasting of storm surges has proven challenging, especially when targeting previously unseen locations or sites without tidal gauges. Furthermore, recent work improved short and medium-term weather forecasting but the handling of raw unassimilated data remains non-trivial. In this paper, we tackle both challenges and demonstrate that neural networks can implicitly assimilate sparse in situ tide gauge data with coarse ocean state reanalysis in order to forecast storm surges. We curate a global dataset to learn and validate the dense prediction of storm surges, building on preceding efforts. Other than prior work limited to known gauges, our approach extends to ungauged sites, paving the way for global storm surge forecasting. 
\end{abstract}

% Note that keywords are not normally used for peerreview papers.
%\begin{IEEEkeywords}
%image reconstruction, multi-modal, multi-temporal, uncertainty.
%\end{IEEEkeywords}

% For peer review papers, you can put extra information on the cover
% page as needed:
% \ifCLASSOPTIONpeerreview
% \begin{center} \bfseries EDICS Category: 3-BBND \end{center}
% \fi
%
% For peerreview papers, this IEEEtran command inserts a page break and
% creates the second title. It will be ignored for other modes.
%\IEEEpeerreviewmaketitle

\section{Introduction}

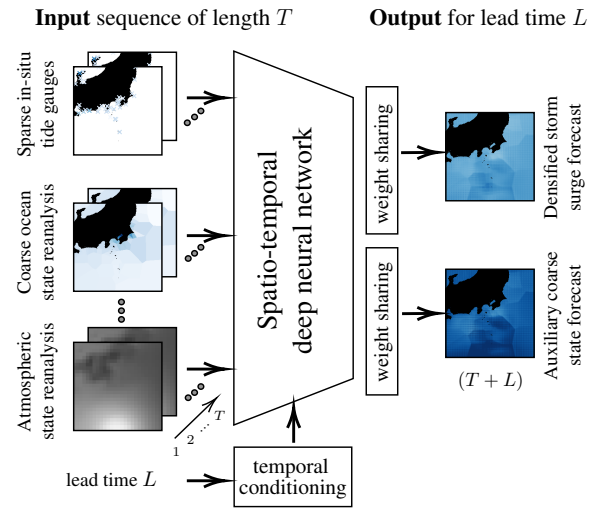
\begin{figure}[ht!]
    \centering
    \resizebox{0.94\linewidth}{!}{%
        \input{Fig/figure_1}
    }
    \caption{\textbf{Overview:} Our approach provides densified high-resolution storm surge forecasts (top) by implicitly assimilating inputs of sparse in situ tide gauge time series (top) with paired sequences of ocean (center) and weather state (bottom) re-analysis products. For additional supervision, coarse ocean state reanalysis maps (bottom) at coinciding lead time are also predicted. 
    }
    \vspace{-2mm}
    \label{fig:teaser}
\end{figure}

Space-borne Earth observation allows for large-scale monitoring of our planet, its atmosphere and events such as natural hazards that may pose significant threat to human life. While the strengths of satellite imagery are its broad spatial extent, its spatio-temporal resolution is inferior compared to on-site measurements. In contrast, in situ sensors may provide (sub-)hourly recordings at highest accuracy, yet they are sparsely deployed and thus lack spatial coverage. Fusing both kinds of data at a global scale holds promises, but harmonizing the sensors in a manner suitable for neural networks to process is an open research direction. A well-established paradigm tackling this issue in the context of weather analysis is that of data assimilation \cite{kalnay2003atmospheric, hersbach2020era5}. However, it is computationally costly and not easily approachable. In this work, we address the challenge of global storm 
\noindent
surge forecasting by implicitly assimilating sparse and raw tide gauge data with coarse weather and ocean state reanalysis products. Storm surges are extreme weather-driven ocean dynamics superimposed on the mean sea level and tidal rhythms, which can cause coastal floods. Scientific consensus is that the coming decades will bring a sharp increase of coastal hazards due to climate change-caused rise in mean sea levels \cite{kirezci2020projections, magnan2022sea, taherkhani2020sea}, aggravated by land subsidence \cite{wang2018effects} and more intense extreme storm events \cite{gori2022tropical, bevacqua2020more}. Our work on storm surge forecasting is motivated to address such hazards, aligning with the \textit{United Nation Sustainable Development Goals} 11.5 \& 13 \cite{sachs2022sustainable, persello2022deep}. Particularly, our approach is inspired by recent advances in weather forecasting that allow generalization to previously unencountered or ungauged sites, which may especially benefit under-served communities with less access to well-maintained in situ measurement infrastructure.
To empower worldwide AI-driven surge forecasting, we curate a novel global and multiple decades spanning dataset of in situ tidal gauge records, paired with weather and ocean state reanalysis, all preprocessed according to the best domain-specific practices.
We highlight the dataset's worth by benchmarking a diverse landscape of approaches including conventional forecasting techniques, an operational numerical model, state-of-the-art deep neural networks, a recent vision transformer for weather forecasting and our enhanced adaptation of a popular lightweight temporal attention network. While we evince the competitiveness of the latter model, the main objective of our experiments is to demonstrate the predictability of storm surges at previously unencountered or altogether ungauged locations, an aim whose feasibility has been questioned in prior work \cite{bruneau2020estimation, tiggeloven2021exploring}.
\medskip

\noindent
In sum, our main contributions are two-fold: 
\begin{itemize}
    \item We introduce a novel, global and multi-decadal dataset of in situ ocean surge time series, paired with atmospheric and ocean state reanalysis products, spuring and facilitating further research on this critical matter.
    \item We demonstrate that precise and dense storm surge forecasts can be obtained by fusing sparse in situ data of coastal tide gauges with coarse atmosphere and ocean reanalysis. Critically, our forecasts extend to previously unseen gauges and entirely ungauged locations, which may benefit under-served communities.
\end{itemize}

\begin{figure*}[ht!]
    \includegraphics[width=\linewidth]{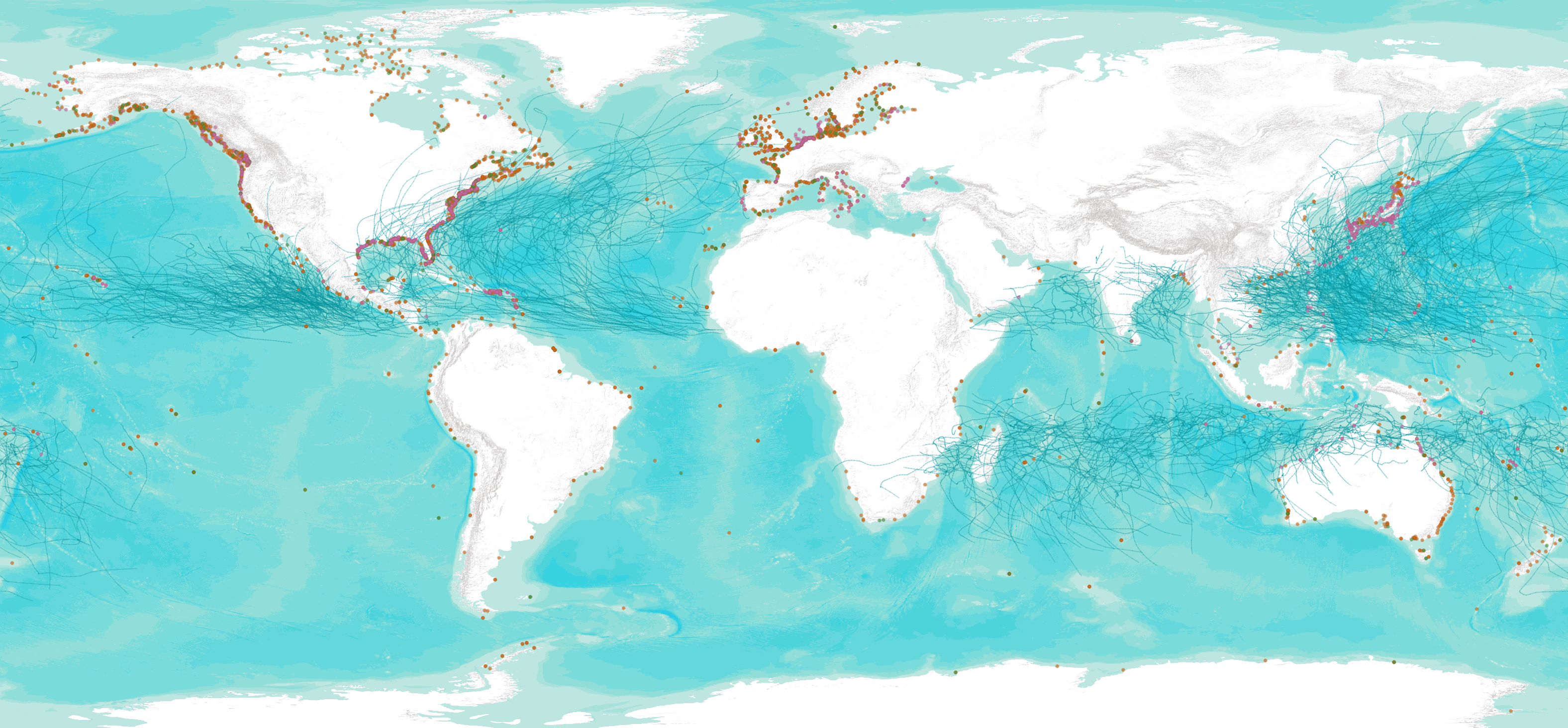}
    \caption{ \textbf{Data.} Green \& orange dots denote storm surge time series locations with records in 1979-2019, as pre-processed from the GESLA-3 collection of tide gauges \cite{haigh2023gesla}. Dark lines indicate hurricane tracks in 2014-2019 as indexed by IBTRaCS \cite{knapp2010international}. Pink markers highlight test split gauges, biased to points of landfall. Visualizations of the ERA5 grid and the irregular GTSM grid are omitted for brevity. 
    }
    \label{fig:map}
\end{figure*}

\section{Related Work}
\subsection{Short-to-medium Range Weather Forecasting}
\label{sec:related:weather}
\noindent
Recent work  marked notable progress on numerical weather prediction \cite{sonderby2020metnet, pathak2022fourcastnet, bi2023accurate, lam2023learning}. 
Such models typically rely on atmospheric initial values from a reanalysis product such as ERA5 \cite{hersbach2020era5}, i.e. best estimates derived by updating prior knowledge with multi-source weather observations.
Contrarily, the contribution of \cite{andrychowicz2023deep} proposes an implicit assimilation approach to fuse in situ weather radar station data with coarse resolution reanalysis products, yielding dense and skillful rainfall forecasts over the United States. While we draw inspiration from this approach, our focus is on the global coastlines to model marine dynamics. 
Technically, our approach deviates from the aforementioned ones by processing local patches of data instead of a coarsely resolved global context in a single forward pass of the network, and is thus significantly more lightweight.

\subsection{Storm Surge Forecasting}
\label{sec:related:surge}

Operational storm surge forecasting pre-dates deep learning, with early techniques explicitly modeling the physics of maritime dynamics with a focus on particular ocean basins \cite{madec2008nemo, sotillo2015myocean}. Of particular interest is the Global Tide and Surge Model (GTSM), a hydrodynamic model forced with ERA5 to globally predict surge on an irregular grid. In this study, we built upon coarsely resolved GTSM ocean state analysis \cite{c3sGTSM} to drive the assimilation of raw in situ data. That is, we fuse the reanalysis product with accurate but sparse in situ records for improved and densified surge predictions. 

Initial efforts for global storm surge modeling via deep learning are given by \cite{bruneau2020estimation, tiggeloven2021exploring}. Both studies are limited to temporal generalization, i.e. they evaluate on gauges trained upon and solely generalize to future time points of these gauges. In contrast, our data and approach enable to generalize in both time and space: Relatedly, recent work \cite{nearing2023ai} proved the feasibility of river streamflow predictions at ungauged basins, in the spirit of which we generalize coastal storm surge forecasting to unseen shores. This defeats the prevailing wisdom that ocean modeling necessitates at least 6-7 prior years of training data at any site of interest \cite{bruneau2020estimation, tiggeloven2021exploring}.

\section{Data} \label{sec:data}
\noindent
We collect a new global multi-decadal dataset combining co-registered atmosphere reanalysis, ocean state reanalysis and pre-processed in situ tide gauge measurements. All data is sampled to an hourly frequency with dates ranging from the beginning of $1979$ to the start of $2019$, and gridded at $0.025 ^{\circ}$ spatial resolution. The atmosphere reanalysis denotes best estimates of historical mean sea level pressure as well as 10 metre U and V wind components ('msl', 'u10' and 'v10', respectively) at about $30$ km resolution, as provided by the ERA5 catalogue \cite{hersbach2020era5}. The ocean-state model forced by ERA5 meteorology provides the storm surge residual based upon the irregularly gridded Deltares Global Tide and Surge Model (GTSM) forced via the aforementioned ERA5 inputs, as given by the Copernicus Climate Change Service (C3S) Climate Data Store (CDS) \cite{c3sGTSM}. Furthermore, a global land-sea mask \cite{zanaga2022esa, kindermann_map} resampled to circa $3$ km resolution is provided. Finally, precise storm surge measurements are derived from in situ tide gauge records collected in GESLA-3 \cite{haigh2023gesla} and spatially distributed as shown in \figref{fig:map}. Principally, our pre-processing pipeline follows the established workflow of \cite{tiggeloven2021exploring}; featuring mean sea level de-trending, harmonic decompositioning \cite{codiga2011unified} and de-noising steps. Key differences to the prior work are their filtering of any in situ sites with records shorter than seven years, whereas we take inspiration from recent progress on (un)gauged river streamflow forecasting \cite{nearing2023ai} and keep such data. While shorter durations pose a greater challenge to learn site-wise dynamics, this drastically increases the overall amount of valuable in situ data from a total of $736$ tidal gauges in \cite{tiggeloven2021exploring} to $3553$ locations in our work --- yielding an almost five-fold increase of valuable in situ data compared to preceding efforts.

\section{Methods}
\noindent
The problem tackled herein is that of forecasting the highly non-linear dynamics of storm surges on a short lead time: Every sample $i=0,1,...,\left|\mathcal{D}\right|$ of the dataset $\mathcal{D}$ denotes a pair $(\boldsymbol{X^i},Y^i)$, with $\boldsymbol{X}^i=[X^i_1, \cdots, X^i_T]$ being the input time series of size $\left[T\times C_{in} \times H \times W \right]$ featuring in situ plus atmospheric reanalysis and model-based surge data, and $Y^i$ is the target image of shape $\left[ C_{out} \times H \times W \right]$ at a lead time of $L$ hours. $T$ is the temporal length of the input series, $C_{in}$ and $C_{out}$ denote the number of input and output channels, and $H\times W$ the images' two spatial dimensions. For convenience, the $i$ superscript is omitted in the remainder of the paper. Unless stated otherwise, we set $T=12$, $C_{in}=5$, $C_{out}=2$, $H=W=256$ and $L=8$ hours, which has been a common choice in prior short-term forecasting works \cite{sonderby2020metnet}. Example data are illustrated in \figref{fig:in_out_plots}, showcasing the time series dynamics, the diversity in context and the presence of extreme weather events in the data.

\subsection{Models}\label{sub:models}
\noindent
We demonstrate the feasibility of our implicit assimilation and densification approach by adapting a representative variety of models. Besides highlighting our paradigm's effectiveness for sparse coastal observations, these evaluations may serve as a benchmark for future research. For each considered network, we follow the respective architecture's best practice in terms of hyperparameters given in the referenced literature, unless otherwise specified. Models are:

\begin{figure*}[!htb]
    \centering
    \begin{subfigure}[b]{0.2166\linewidth}
        \includegraphics[width=\linewidth]{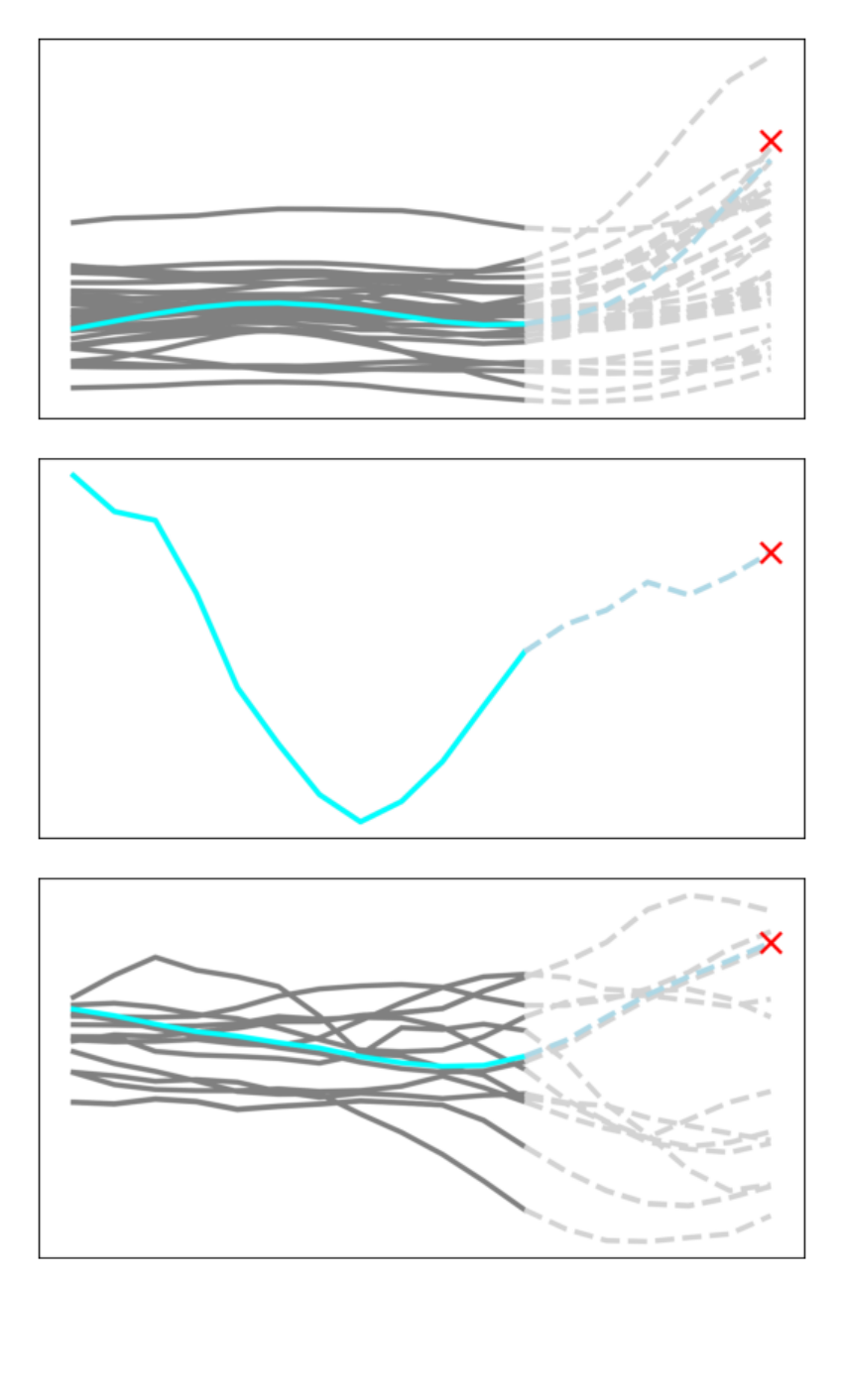}
        \caption{In situ time series.}
    \end{subfigure}
    \begin{subfigure}[b]{0.5482\linewidth}
        \includegraphics[width=\linewidth]{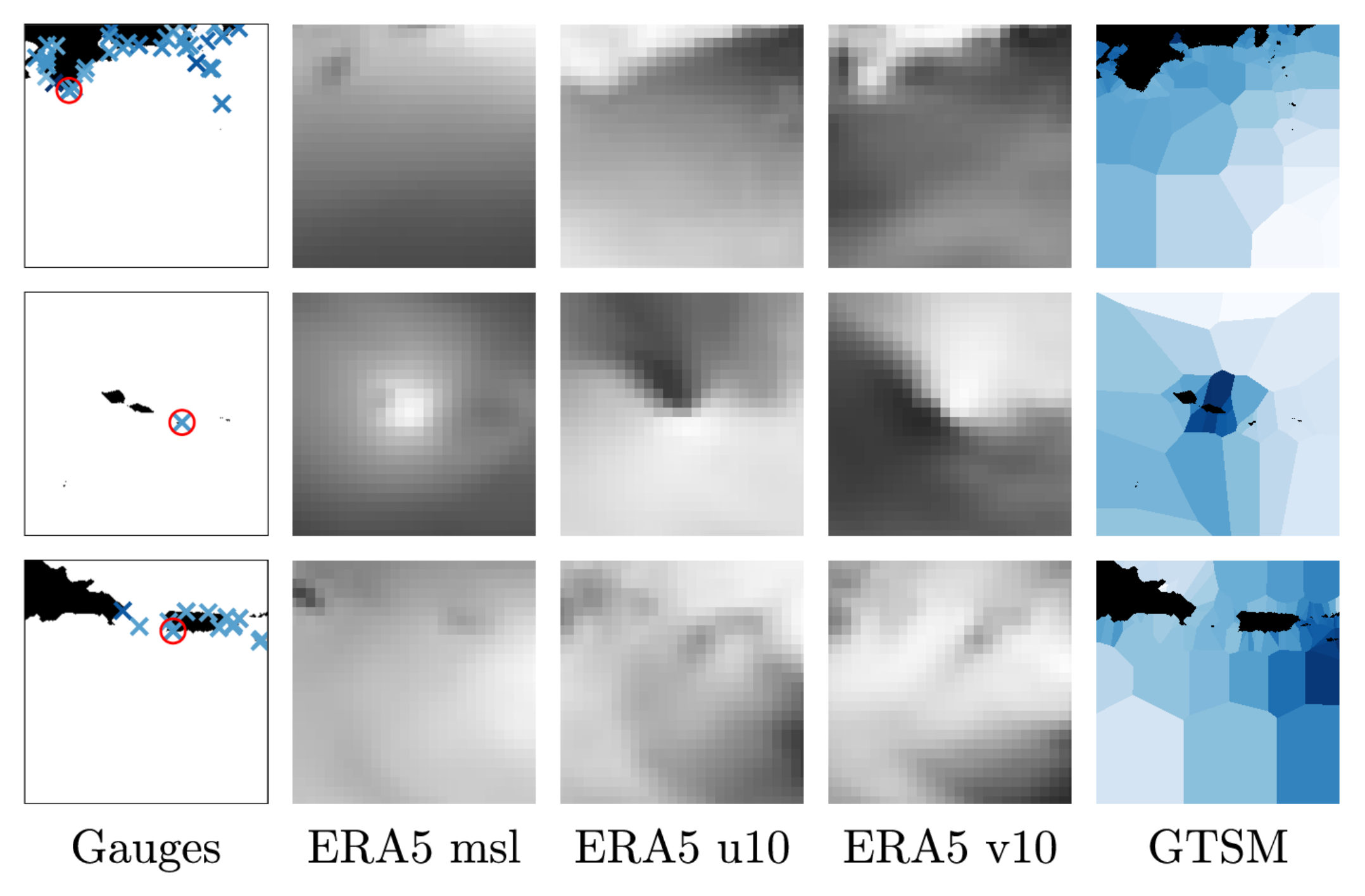}
        \caption{Input}
    \end{subfigure}
    \begin{subfigure}[b]{0.2252\linewidth}
        \includegraphics[width=\linewidth]{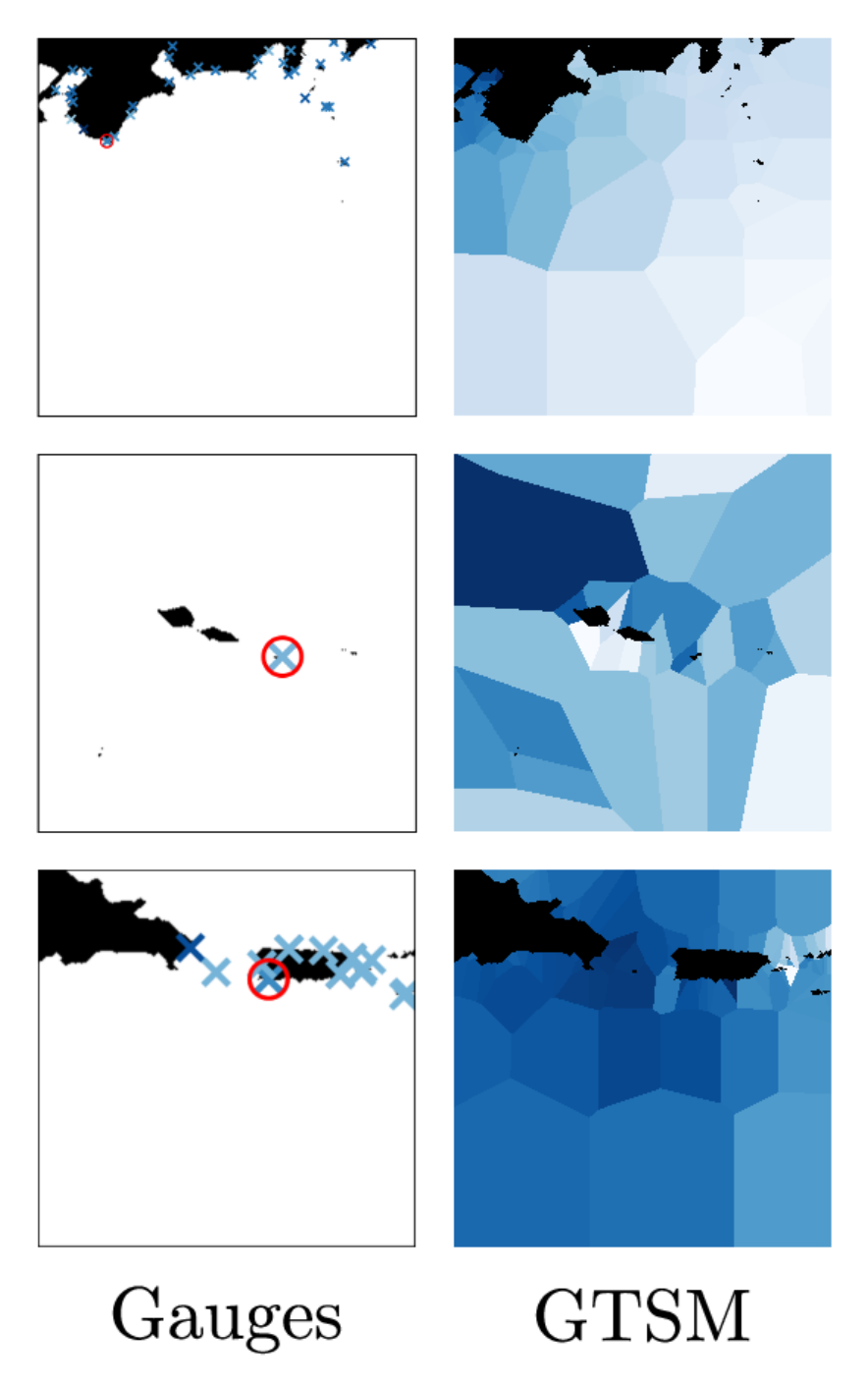}
        \caption{Target}
    \end{subfigure}
    
  \caption{\textbf{Example data}, one local sample per row. Input grouped at the left, targets at the right. Inputs: Time series of target (blue) and context gauges (grey), with target history only given in the hyperlocal setting. gauge locations and surge values. ERA5 pressure at mean sea level, wind at 10 m in u and v directions. Coarse GTSM input. Targets: Surge at target time, GTSM at target time.
  }
  \label{fig:in_out_plots}
\end{figure*}

\paragraph{\bf Conventional baselines} As a simple baseline, we consider the seasonal average surge based on historic values at the gauge of interest and the given target time. This necessitates historical data at the target gauge, but is expected to provide a solid baseline for sites experiencing seasonal cyclone activity \cite{shan2023seasonal}. Second, we consider the mean surge at the gauge of interest over the input time period. While again requiring access to the gauge's records, this would prove beneficial whenever the surge at target time doesn't deviate too much from the input period's. Third, we consider the linear extrapolation of surge time series inputs to the target time. Finally, the global physical storm surge predictions of GTSM forced with ERA5 data are reported and compared against \cite{c3sGTSM}. As GTSM is numerically simulated on an irregular grid \cite{kernkamp2011efficient, deltares2018delft3d} we perform nearest neighbor extrapolation to get coarse forecasts at any site of interest, specifically for getting predictions at unknown test split gauges, and then extrapolate these to the target time.

\vspace{-1.75mm}
\paragraph{\bf LSTM-based} Long short-term memory (LSTM) networks \cite{hochreiter1997long} are the state-of-the-art architectures for temporal modeling of fluvial \cite{kratzert2018rainfall, nearing2023ai} and coastal dynamics \cite{tiggeloven2021exploring}. Accordingly, we consider classical as well as convolutional LSTM (ConvLSTM) \cite{shi2015convolutional} for storm surge forecasting. We use the architectures from \cite{tiggeloven2021exploring} and minimally adapt them for the sake of comparability to accept our more comprehensive data, including time series of preceding gauge values. As such, both models receive a 5x5 context window around the gauge, and predict a single point in the future. 

\paragraph{\bf Attention-based} We consider spatio-temporal transformer models, which both principally share a common structure of inputs and outputs as depicted in \figref{fig:teaser}. First, we evaluate a MaxVIT U-Net backbone \cite{tu2022maxvit} as recently proposed for weather forecasting in \cite{andrychowicz2023deep}, adapted to our problem statement. The network collates temporal information into the channel dimension, but is conditioned on the lead time of the target via Feature-wise Linear Modulation (FiLM) \cite{perez2018film}. 
Finally, we consider the U-TAE of \cite{garnot2021panoptic}, originally proposed for panoptic segmentation. We adjust U-TAE by introducing FiLM at each of its convolution blocks---such that additionally to temporal embeddings at the input time points, our adaptation of the model is conditioned to forecast at a variable target time. Notably, the key difference between the last two models is that \cite{tu2022maxvit} resolves the input time series into the channel dimension and doesn't model time dynamics explicitly, whereas \cite{garnot2021panoptic} processes time series explicitly and applies lightweight temporal attention but doesn't explicitly model global spatial interactions. 

\subsection{Densification}
\noindent
Central to our approach generalizing storm surge forecasting to previously unseen or ungauged sites is the concept of densification. For any model that outputs a two-dimensional storm surge forecast map $\boldsymbol{\hat{y}}_{s}$, we implement densification via the built-in spatial parameter sharing of the convolution operator. Specifically, we utilize 1 $\times$ 1 convolution kernels with a stride of $1$ at the final network layer to broadcast from sparsely populated to non-observed pixel coordinates in the spatial dimensions. Auxiliary supervision and input data dropout are also used to further encourage the networks to learn densification, as proposed by Andrychowicz et al \cite{andrychowicz2023deep}.

\paragraph{\bf Auxiliary supervision}
Complementarily to learning a densified forecast of the sparse in situ time series, the networks additionally predict a forecast $\boldsymbol{\hat{y}}_{c}$ of the coarse GTSM ocean state at the same lead time $L$, as depicted in \figref{fig:teaser}. This way, the models receive additional feedback at pixels which would otherwise not be populated and the preceding shared layers internalize to implicitly assimilate the sparse observations with the coarse reanalysis. To only evaluate the coarse ocean state predictions over valid locations we mask the loss computation with a land sea mask $\mathds{1}_{lsm}$.

\vspace{-2.45mm}
\paragraph{\bf In situ dropout} To furthermore encourage the densifying networks to predict non-trivial outputs at unpopulated pixels within the sparse input time series we perform data dropout. Specifically, we randomly remove in situ tide gauges from the input with a probability $p$ but keep all sites within the target patch, such that the network is forced to learn extrapolating to the dropped sites. We set $p=0.25$ and include a binary validity mask $\mathds{1}_{val}$ in the network inputs as proposed in a weather prediction context by \cite{andrychowicz2023deep}. \newline

\noindent
In sum, the densifying network architectures output two maps $\boldsymbol{\hat{y}} = [\boldsymbol{\hat{y}}_{s}, \boldsymbol{\hat{y}}_{c}]$. Map $\boldsymbol{\hat{y}}_{s}$ densely predicts the sparse GESLA gauges, and $\boldsymbol{\hat{y}}_{c}$ predicts the spatially interpolated future coarse GTSM values. Thus, they are trained via a weighted combination of two masked L1 cost functions

\vspace{-0.35mm}
\begin{equation}
\mathcal{L}_{s}(\boldsymbol{\hat{y}}_{s},\boldsymbol{y}_{GESLA}) = \frac{1}{n} \sum_{j=1}^{n}  \mathds{1}_{val}(j) \cdot \|\boldsymbol{\hat{y}}_j - \boldsymbol{y}_j\|_1 \; ,
\label{eq:L2}
\end{equation}

\vspace{-1.6mm}
\begin{equation}
\mathcal{L}_{c}(\boldsymbol{\hat{y}}_{c},\boldsymbol{y}_{GTSM}) = \frac{1}{n} \sum_{j=1}^{n}  \mathds{1}_{lsm}(j) \cdot \|\boldsymbol{\hat{y}}_j - \boldsymbol{y}_j\|_1 \; ,
\label{eq:L2}
\end{equation}

masked via $\mathds{1}_{val}$ and $\mathds{1}_{lsm}$, resulting in the combined loss

\begin{equation}
\mathcal{L}(\boldsymbol{\hat{y}},\boldsymbol{y}) = \mathcal{L}_{s}(\boldsymbol{\hat{y}}_{s},\boldsymbol{y}_{GESLA}) + \lambda \mathcal{L}_{c}(\boldsymbol{\hat{y}}_{c},\boldsymbol{y}_{GTSM}) \; .
\label{eq:L2}
\end{equation}

We set the hyperparameter $\lambda = \frac{1}{100}$, to account for the sparseness of in situ data in comparison to the pixel-wise coarse evaluations and to compensate for the resulting magnitudes of differences in supervision frequency across both domains as well as their respective loss terms.

\subsection{Lead time conditioning}
\label{sub:lead_time}
\noindent
To enable a flexible forecasting, accommodating for varying hours of look-ahead predictions at inference time and via a single forward-pass, we implement lead time conditioning via Feature-wise Linear Modulation (FiLM) \cite{perez2018film}. Specifically, a shallow encoder projects queried lead times $L$ into a low-dimensional feature space and linearly modulates convolutional feature maps via a learned scale and bias offset. We utilize lead time conditioning for the two considered densifying networks, and have one shallow encoder per each of their U-Net backbone's convolutional blocks.

\section{Experiments}

\paragraph{Splits} For our experiments, we set up splits by defining holdout data in terms of both the spatial and temporal dimensions: A globally distributed 20 \% of coastal gauges across all ocean basins are reserved for the test split, whose temporal extent starts from April $2014$ and is thus well observed via satellites, in the interest of follow-up works. The remaining 70 and 10 \% of locations are utilized as training and validation splits, with records ranging from $1979$ to $2014$. This amounts to a total of $2561$, $284$ and $708$ gauges for our train, validation and test split, respectively.
The resulting map of in situ recordings (color-coded according to their splits) and hurricane trajectories is depicted in Fig. \ref{fig:map}. The distribution of accessible and gauged sites is biased towards developed countries, underlining the need for machine learning solutions to serve under-represented regions.

\paragraph{Sampling} Having assigned a subset of gauges to the holdout splits, we determine the test dates by identifying coincidences of in situ records with storm tracks as given by IBTrACS \cite{knapp2010international}. When a storm passes within $100$ km of a holdout tide gauge then we set such a date as the sample's target time. If no storm tracks pass by, then we resort to sampling target times at outlier surge values deviating more than 2 train split standard deviations from the train split's mean surge. At train time, we likewise perform outlier sampling with a probability of $0.5$, roughly mirroring the distribution of (non) storm events at holdout gauges in the validation and test splits. If no such sample exists for the current gauge, then a random target date is drawn instead. 

\begin{figure}[!h] 
    \resizebox{0.96\linewidth}{!}{%
        \input{Fig/exp_setup.tex}
    }
    \caption{\textbf{Experimental setup.} Design of the densification and hyperlocal evaluation schemes, conceptualizing their respective inputs and outputs. The hyperlocal protocol focuses on forecasting of novel dynamics encountered at inference time, predicting surge at holdout target (green) and context gauges (blue) $L$ hours ahead. The generalization setup quantifies the goodness of models to broadcast predictions to ungauged locations, i.e. unknown gauges not contained in the input and solely used for evaluation.
    }
    \label{fig:exp_setup}
\end{figure}
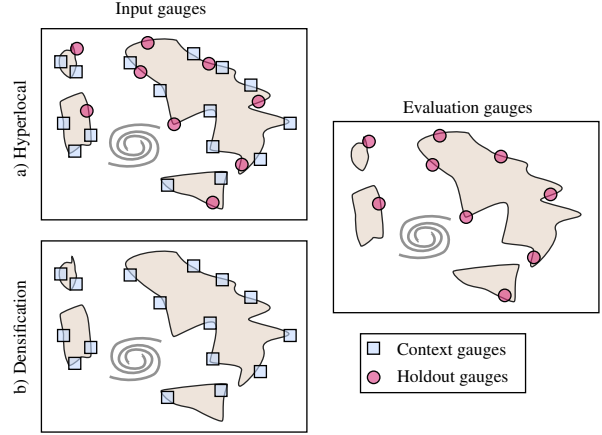

\subsection{Implementation details}
\noindent
To enable fast online sampling of data and efficient training of deep networks, we represent all spatio-temporal data as netCDF files \cite{rew1990netcdf} via xarray \cite{hoyer2017xarray}, either loaded directly into memory or read in parallel via dask \cite{rocklin2015dask}. This is especially critical for the in situ GESLA-3 records, which are re-processed into a compact format as part of the preprocessing pipeline and get released with this publication.
\normalsize

In contrast to weather forecasting models \cite{pathak2022fourcastnet, bi2023accurate, lam2023learning} that process a global context window at proximately $30$ km resolution, the networks we consider are significantly less resource-demanding and digest local patches of ($256$ px)$^2$ at a finer pixel-resolution of circa $3$ km to capture local variations of surge. All data features are z-standardized via their sufficient statistics calculated on the training split.

\paragraph{\bf Training}

For training, 1 epoch is defined by iterating over all train split gauges in a random order. At each gauge, a Gaussian is drawn around it's location to randomly sample what is treated as the local patches centroid $c$. The target date \& time $t$ are drawn randomly from the current gauge's records. Next, a lead time $L$ is drawn randomly from $\{0,1,2,...,12\}$, as in \cite{andrychowicz2023deep}. The hourly input time sequence of length $T$ is then given by the time interval of $[t-L-T, t-L]$. Note that the random sampling of $c$, $t$ and $L$ effectively acts as data augmentation. For auto-regressive methods we evaluate the prediction at time $t$, whereas single-forward-pass approaches based on FiLM are conditioned on $L$ to directly generate a forecast at time $t$. 

\begin{table*}[!t]
\centering
\caption{\textbf{Experimental evaluation.} We benchmark forecasting skills for $T=12$ and $L=8$ on a global holdout set of previously unseen gauges. Point-wise storm surge predictions at known gauges are evaluated in the hyperlocal setting (left), whereas the densification protocol (right) tests predictions at unknown or ungauged sites. FiLM U-TAE is the most competitive approach, followed by MaxVIT U-Net.}
\scalebox{0.90}{
\begin{tabular}{@{}lllllllll@{}}
\toprule
\textbf{Model} & \multicolumn{3}{c}{\textbf{Hyperlocal}} & \multicolumn{3}{c}{\textbf{Densification}} \\ \cmidrule(lr){2-4} \cmidrule(lr){5-7}
        & $\downarrow$ MAE (std) & $\downarrow$ MSE (std) & $\uparrow$ NNSE  & $\downarrow$ MAE (std) & $\downarrow$ MSE (std) & $\uparrow$ NNSE \\ \midrule
seasonal average & 0.281 (0.313)        & 0.177 (0.539) & 0.424 & ---        & --- & ---\\
input average & 0.267 (0.295)        & 0.158 (0.452) & 0.452 & ---        & --- & --- \\
input extrapolation & 0.182 (0.239) &  0.090 (0.342) &  0.593 & --- &  --- &  --- \\
GTSM extrapolation \cite{c3sGTSM} & --- &  --- &  --- & 0.351 (0.643) &  0.536 (4.744) & 0.195 \\
\midrule
LSTM \cite{hochreiter1997long, tiggeloven2021exploring}   & 0.166 (0.282) & 0.107 (0.759) & 0.595 & ---       & --- & --- \\
ConvLSTM \cite{shi2015convolutional, tiggeloven2021exploring}   & 0.162 (0.267) & 0.098 (0.691) & 0.618 & ---        & --- & --- \\
\midrule
FiLM U-TAE \cite{perez2018film, garnot2021panoptic}  & \textbf{0.158 (0.209)}        & \textbf{0.069 (0.248)} & \textbf{0.655} & 0.190 (0.260)        & \textbf{0.104 (0.535)} & \textbf{0.556} \\
MaxVIT U-Net \cite{tu2022maxvit, andrychowicz2023deep}   & 0.160 (0.212)        & 0.070 (0.263) & 0.649 & \textbf{0.178 (0.273)}        & 0.106 (0.587) & 0.552 \\
\bottomrule
\end{tabular}
 }
\newline
\vspace*{.05 cm}
    \label{tab:main_results}
\end{table*}

We use the ADAM optimizer \cite{kingma2014adam} at a batch size of 16, with initial learning rates tuned over magnitudes $10^{-1}$ to $10^{-4}$ for each model individually. All networks train for 50 epochs with an exponential learning rate decay of 0.9. Models are evaluated on the validation split each epoch and the checkpoint with best validation loss is used for testing. 
\label{sub:training}

\subsection{Evaluation}
\noindent 
All network predictions at target time are compared against their respective test split in situ tide gauge values. Prediction goodness is evaluated in terms of Mean Absolute Error (MAE) as well as Mean Squared Error (MSE), reported in units of meters and with error-wise standard deviations (std) across the set of test split gauges denoted in brackets. Finally, we report each method's Normalized Nash-Sutcliffe Efficiency (NNSE) \cite{nash1970river}, which conceptually relates to the coefficient of determination ($R^2$) and takes values within $0$ and $1$.
Similar to prior weather forecasting work \cite{andrychowicz2023deep} we evaluate in two experimental setups. The concepts of both setups are depicted in \figref{fig:exp_setup} and given as follows:

\paragraph{\bf i. Hyperlocal evaluation} In this experimental paradigm, the time series of holdout gauges not trained upon are included into the model's inputs at test time. Therefore, the challenge becomes to integrate previously unseen dynamics at novel locations and to assimilate newly encountered tidal gauges at inference time.

\paragraph{\bf ii. Densification evaluation} Predictions at previously unseen test split gauges are obtained via densification, i.e. a model's ability to predict at unknown locations is quantified here. Importantly, test split gauges are not part of the input time series and only used as targets. Note that this setup can't be accomplished via previous established approaches for storm surge forecasting not implementing densification, e.g. the conventional baselines and LSTM-based models. 

\section{Results}

\subsection{Main experiments}
\noindent
To demonstrate the feasibility of our problem statement and the benefits of our curated dataset, we evaluate all considered approaches according to their applicability in the hyperlocal and densification experimental schemes. Outcomes are reported in \tabref{tab:main_results}. The results show that FiLM U-TAE performs best in the hyperlocal setting, forecasting surge at newly encountered gauges with a mean absolute accuracy of circa $16$ cm. The seasonal average and input average predictions are more erroneous, particularly in terms of squared error---validating the presence of storm-driven extremes in the test data. Overall, all neural networks outperform competing approaches in the hyperlocal setting. 

In the densification experiment, FiLM U-TAE performs best, closely followed by MaxVIT U-Net. Notably, both densifying models denote a substantial improvement over the GTSM baseline, whose prediction goodness exhibits elevated standard deviations across gauges. 

\begin{figure*}[!h] 
    \centering
    \includegraphics[width=0.85\linewidth]{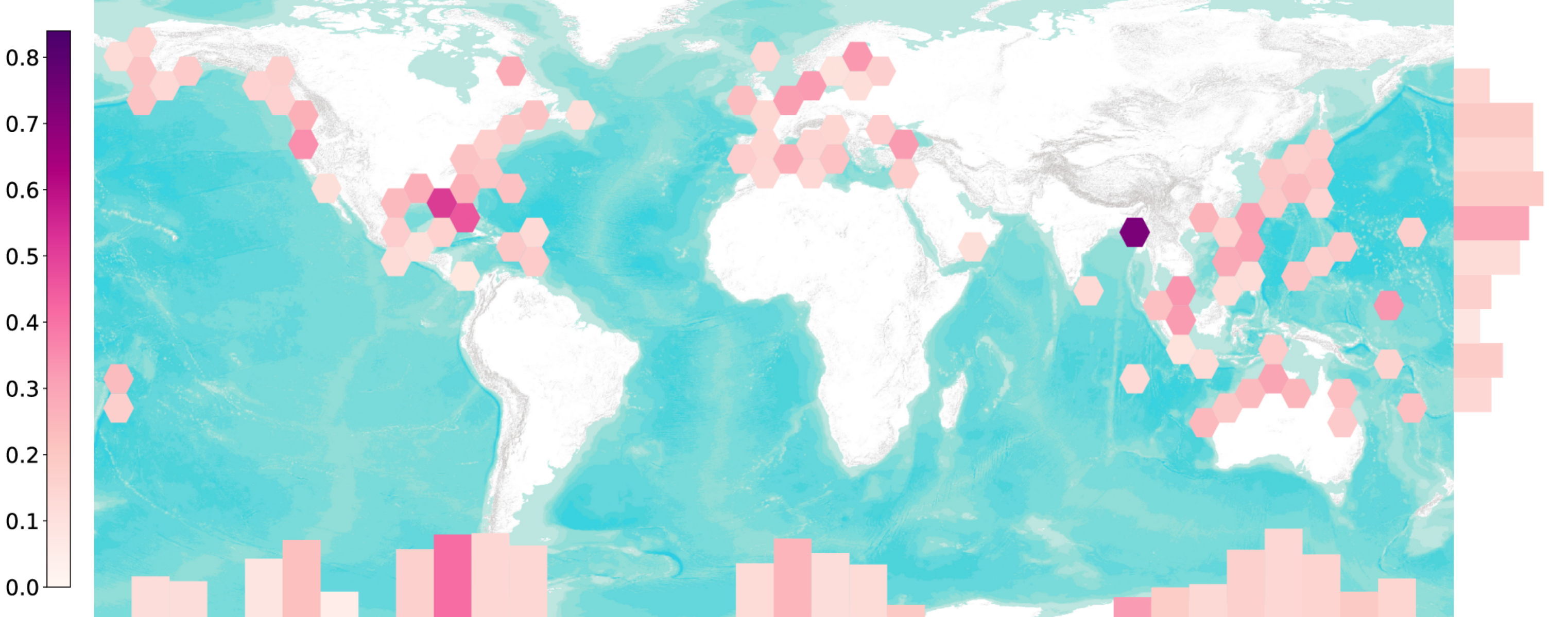}
    \caption{\textbf{Location} of holdout gauges impacts prediction performance. Average absolute errors in meters are color-coded and binned according to each sites' longitude and latitude coordinates, with gauge counts overlayed for each spatial dimension. Particularly challenging regions are the Gulf of Mexico, the Caribbean Sea and the Indian Ocean due to their extreme climate and resulting outsized surge dynamics.
    }
    \label{fig:spatial_depend}
\end{figure*}

Altogether, FiLM U-TAE tends to outperform MaxVIT U-Net, implying that temporal self-attention is of greater benefit than visual attention for our spatio-temporal forecasting task. To convey a better understanding of the spatial dependency of errors, \figref{fig:spatial_depend} analyzes our best model's performance across the globe. The analysis shows that forecasting in the tropics is particularly hard, confirming that the curated benchmark provides a challenging problem.

\subsection{Ablation experiments}
\noindent
In order to further investigate the challenges of short-term storm surge forecasting and explore which design choices determine the quality of predictions, we conduct a series of ablation studies following the densification experimental protocol. All ablations are run with the FiLM U-TAE network, which the preceding main experiments identified as the best performing backbone for our approach.

\begin{table}[!ht]
\caption{\textbf{Repeated Measures.} Evaluation of FiLM U-TAE with varying numbers of input time points $T$, flexibly accommodated for via temporal self-attention. Longer inputs tend to be beneficial.}
\centering
\scalebox{0.76}{
\begin{tabular}{@{}llllllllll@{}}
\toprule
& input length T & { $\downarrow$ MAE (std)} & {$\downarrow$ MSE (std)} & $\uparrow$ NNSE \\ \midrule
& 6 & 0.194 (0.282) & 0.115 (0.587) & 0.551 \\
& 12 & 0.190 (0.260) & 0.104 (0.535) & 0.556 \\
& 18 & \textbf{0.180 (0.230)} & \textbf{0.085 (0.510)} & \textbf{0.573} \\
& 24 & \textbf{0.180 (0.230)} & \textbf{0.085 (0.510)} & 0.571 \\
\bottomrule
\end{tabular}
}
    \label{tab:input_len}
\end{table}

\begin{table}[!ht]
\caption{\textbf{Lead Time.} Evaluation of FiLM U-TAE with varying lead time offset $L$, modifiable thanks to lead time conditioning. Storm surge forecasts become more challenging the larger $L$ gets.
    }
\centering
\scalebox{0.76}{
\begin{tabular}{@{}llllllllll@{}}
\toprule
& lead time t & { $\downarrow$ MAE (std)} & {$\downarrow$ MSE (std)} & $\uparrow$ NNSE \\ \midrule
& 4 & \textbf{0.169 (0.254)} & \textbf{0.093 (0.543)} & \textbf{0.583} \\
& 6 & 0.182 (0.269) & 0.106 (0.551) & 0.552 \\
& 8 & 0.190 (0.260) & 0.104 (0.535) & 0.556 \\
& 10 & 0.191 (0.273) & 0.111 (0.553) & 0.540 \\
& 12 & 0.196 (0.273) & 0.113 (0.539) & 0.536 \\
\bottomrule
\end{tabular}
}
    \label{tab:lead_time}
\end{table}

\begin{table}[!ht]
\caption{\textbf{Input ablations.} Evaluation of our models with varying inputs. The outcomes underline the relevance of each modality.}
\centering
\scalebox{0.75}{
\begin{tabular}{@{}lllllllll@{}}
\toprule
& input ablation & { $\downarrow$ MAE (std)} & {$\downarrow$ MSE (std)} & {$\uparrow$ NNSE} \\ \midrule
& full model & 0.190 (0.260) & \textbf{0.104 (0.535)} & \textbf{0.556} \\
& no GTSM input & 0.207 (0.284) & 0.124 (0.543) & 0.513  \\
& no ERA5 input & 0.189 (0.273) & 0.110 (0.545) & 0.542  \\
& no data dropout & 0.217 (0.289) & 0.130 (0.539) & 0.500  \\
& no FiLM, $L=8$ fixed & \textbf{0.183 (0.273)} & 0.108 (0.567) & 0.547  \\ 
\bottomrule
\end{tabular}
}
\label{tab:in_ablations}
\end{table}

\begin{table}[!ht]
\caption{\textbf{Output ablations.} Evaluation of FiLM U-TAE with varying output channels. Ablations show all outputs' significance. 
    }
\centering
\scalebox{0.71}{
\begin{tabular}{@{}lllllllll@{}}
\toprule
& output ablation & { $\downarrow$ MAE (std)} & {$\downarrow$ MSE (std)} & {$\uparrow$ NNSE} \\ \midrule
& full model & \textbf{0.190 (0.260)} & \textbf{0.104 (0.535)} & \textbf{0.556} \\
& no GTSM supervision & 0.194 (0.276) & 0.114 (0.544) & 0.534   \\
& GTSM, instead of densification & 0.210 (0.246) & 0.105 (0.536) & 0.554 \\
\bottomrule
\end{tabular}
}
\label{tab:out_ablations}
\end{table}

\paragraph{\bf Accuracy vs. sequence length} To evaluate the effect of the number of input time points $T$ on performances, we run FiLM U-TAE on input time series of lengths $T = 6, 12, 18, 24$ hours. \tabref{tab:input_len} shows that longer sequences drive better forecasts, although gains saturate. This confirms the intuition that more context and data regarding maritime dynamics facilitates short-term forecasting, but that observations further in the past become less informative. 

\paragraph{\bf Accuracy vs. lead time} To evaluate the effect of the lead time $L$ on performances, we perform inference by varying $L=4,6,8,10,12$ hours ahead. Note that the lead time can be systematically varied in a single forward pass thanks to the network being directly conditioned on $L$. The outcomes in \tabref{tab:lead_time} validate the intuition that longer lead times exacerbate the prediction problem, affirming its non-linear dynamics and the challenges of medium-term forecasting as encountered by numerical weather prediction models. 

\paragraph{\bf Ablation studies} We systematically ablate over input information and supervision extent to investigate each element's significance in driving the prediction goodness. The results are reported in Tables \ref{tab:in_ablations} and \ref{tab:out_ablations}, respectively. Input ablations justify the provisioning of coarse GTSM plus ERA5 auxiliary inputs, and show that the network can effectively translate atmospheric reanalysis to ocean states. Furthermore, data dropout is critical for enabling densification and the introduction of flexible lead time conditioning is beneficial. The output ablations confirm that coarse GTSM supervision provides valuable guidance, yet there is additional gains for the densified storm surge output. In sum, the outcomes validate our overall approach as depicted in \figref{fig:teaser}. 

\begin{figure*}[!htb]
  \centering
  \begin{subfigure}[b]{0.185\linewidth}
    \includegraphics[width=\linewidth]{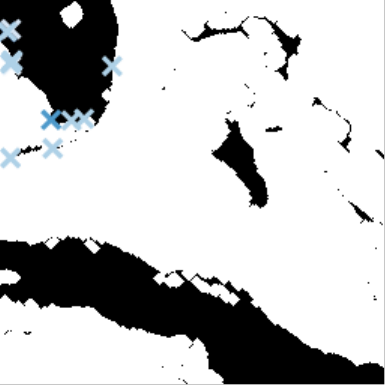}
  \end{subfigure}
  \begin{subfigure}[b]{0.185\linewidth}
    \includegraphics[width=\linewidth]{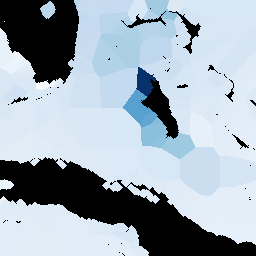}
  \end{subfigure}
  \begin{subfigure}[b]{0.185\linewidth}
    \includegraphics[width=\linewidth]{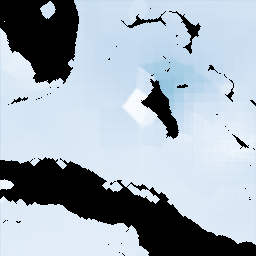}
  \end{subfigure}
  \begin{subfigure}[b]{0.185\linewidth}
    \includegraphics[width=\linewidth]{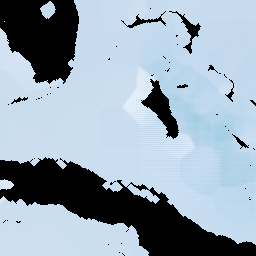}
  \end{subfigure}
  \begin{subfigure}[b]{0.185\linewidth}
    \includegraphics[width=\linewidth]{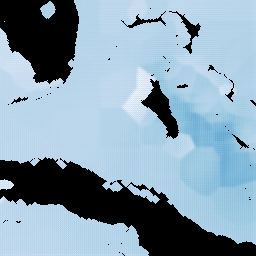}
  \end{subfigure}
  
  \begin{subfigure}[b]{0.185\linewidth}
    \includegraphics[width=\linewidth]{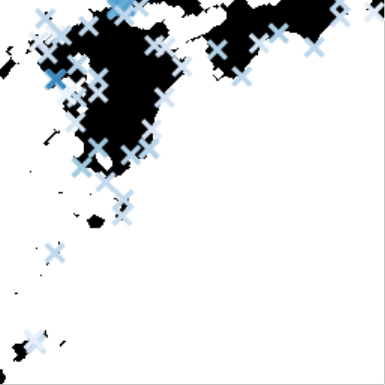}
  \end{subfigure}
  \begin{subfigure}[b]{0.185\linewidth}
    \includegraphics[width=\linewidth]{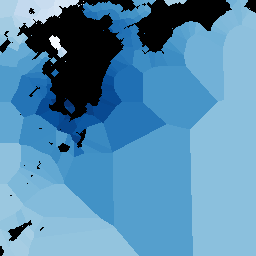}
  \end{subfigure}
  \begin{subfigure}[b]{0.185\linewidth}
    \includegraphics[width=\linewidth]{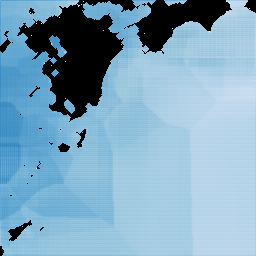}
  \end{subfigure}
  \begin{subfigure}[b]{0.185\linewidth}
    \includegraphics[width=\linewidth]{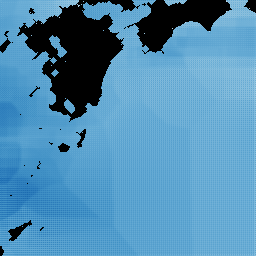}
  \end{subfigure}
  \begin{subfigure}[b]{0.185\linewidth}
    \includegraphics[width=\linewidth]{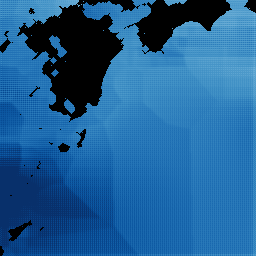}
  \end{subfigure}
  
  \begin{subfigure}[b]{0.185\linewidth}
    \includegraphics[width=\linewidth]{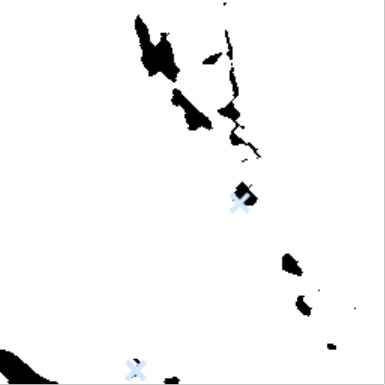}
  \end{subfigure}
  \begin{subfigure}[b]{0.185\linewidth}
    \includegraphics[width=\linewidth]{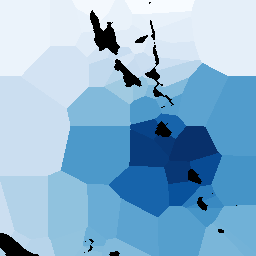}
  \end{subfigure}
  \begin{subfigure}[b]{0.185\linewidth}
    \includegraphics[width=\linewidth]{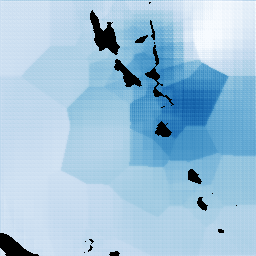}
  \end{subfigure}
  \begin{subfigure}[b]{0.185\linewidth}
    \includegraphics[width=\linewidth]{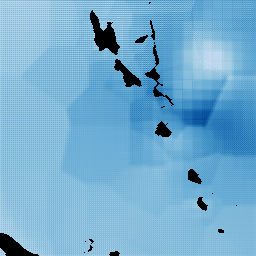}
  \end{subfigure}
  \begin{subfigure}[b]{0.185\linewidth}
    \includegraphics[width=\linewidth]{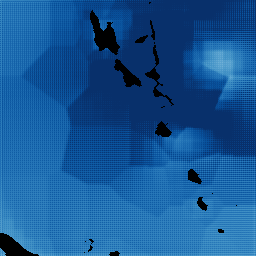}
  \end{subfigure}  
   
  \begin{subfigure}[b]{0.185\linewidth}
    \includegraphics[width=\linewidth]{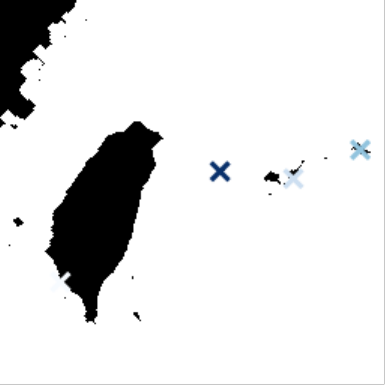}
    \caption{gauge locations.}
  \end{subfigure}
  \begin{subfigure}[b]{0.185\linewidth}
    \includegraphics[width=\linewidth]{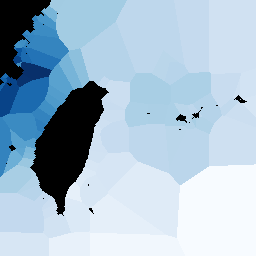}
    \caption{GTSM prediction.}
  \end{subfigure}
  \begin{subfigure}[b]{0.185\linewidth}
    \includegraphics[width=\linewidth]{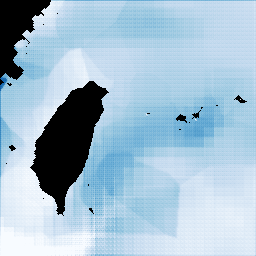}
    \caption{VIT U-Net densification.}
  \end{subfigure}
  \begin{subfigure}[b]{0.185\linewidth}
    \includegraphics[width=\linewidth]{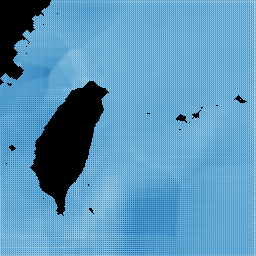}
    \caption{U-TAE densification.}
  \end{subfigure}   
  \begin{subfigure}[b]{0.185\linewidth}
    \includegraphics[width=\linewidth]{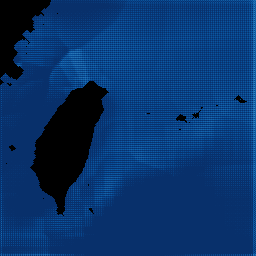}
    \caption{U-TAE coarse prediction.}
  \end{subfigure}

  \caption{\textbf{Exemplary data and densified storm surge forecasts} in the densification experimental setup. Rows: Four samples from the test split. Columns: Sampled gauge locations. Dense surge forecasts of GTSM, MaxVit U-Net, FiLM U-TAE and coarse auxiliary FiLM U-TAE predictions. All illustrated outputs are in the densification setup without the target gauge provided, at a lead time of $L=8$ hours.
  }
  \label{fig:bunch_of_plots}
\end{figure*}

\paragraph{\bf Qualitative results} Complementary to the reported quantitative outcomes, \figref{fig:bunch_of_plots} illustrates example data and the different densification models' forecasts. Notably, the networks' densified predictions show substantial differences from GTSM, as they are driven by the assimilation of in situ tidal gauge data which GTSM does not incorporate. Specifically, modifications are undertaken close to the shorelines where gauges are present and surge forecasts are most relevant. Furthermore, the densifications often differ from the auxiliary coarse predictions, underlining the functional differences across the two kinds of outputs and highlighting once more the importance of integrating the sparse in situ data. Finally, the appearance of gridding in the spatio-temporal networks' outputs evidences the impact of the ERA5 atmospheric information, which is integrated in the storm surge forecasts.

\section{Conclusion}
\noindent
To tackle the aggravating hazard of coastal floods, we introduce a novel dataset and framework forecasting storm surges. Our curated data makes the posed challenge more accessible to the remote sensing community and may serve as a benchmark to fuel future research. Our approach is influenced by recent progress in weather forecasting, and shows that neural networks can implicitly assimilate sparse in situ measurements with coarse weather and ocean state reanalysis products to provide densified forecasts. In a follow-up, we'll explore the operational potential of our approach and replacing retrospective reanalysis products with recently developed forecasting models. Further directions may be the incorporation of satellite altimetry, the modeling of impact at landfall and the translation from storm surges to predicting flood maps. Data and code are given at \small{\url{https://github.com/PatrickESA/StormSurgeCastNet}}.

\clearpage

%%%%%%%%% REFERENCES
{\small
\bibliographystyle{ieee_fullname}
\bibliography{egbib}
}

\end{document}

%% file: Fig/figure_1.tex
\tikzset{every picture/.style={line width=0.75pt}} %set default line width to 0.75pt        

\begin{tikzpicture}[x=0.75pt,y=0.75pt,yscale=-1,xscale=1]
%uncomment if require: \path (0,372); %set diagram left start at 0, and has height of 372

%Image [id:dp447919462677264] 
\draw (330.24,107.36) node  {\includegraphics[width=45.6pt,height=45.6pt]{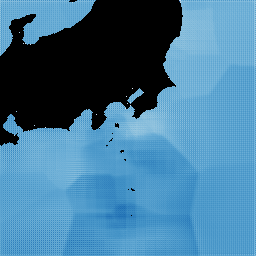}};
%Image [id:dp8969336409330839] 
\draw (330.24,215.36) node  {\includegraphics[width=45.6pt,height=45.6pt]{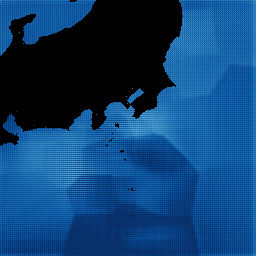}};
%Image [id:dp8806348634582146] 
\draw (83.84,65.76) node  {\includegraphics[width=45.6pt,height=45.6pt]{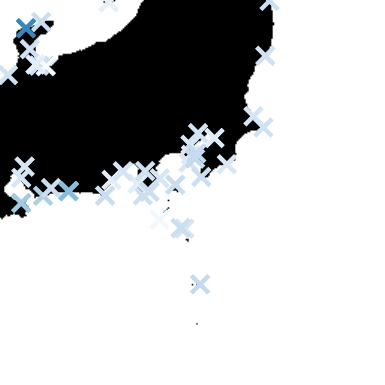}};
%Shape: Rectangle [id:dp5536973774369096] 
\draw   (53.44,35.36) -- (114.24,35.36) -- (114.24,96.16) -- (53.44,96.16) -- cycle ;

%Image [id:dp8377017738175867] 
\draw (73.84,75.76) node  {\includegraphics[width=45.6pt,height=45.6pt]{Fig/bunch/3/4_in_scatter_lsm.pdf}};
%Shape: Rectangle [id:dp664568679343208] 
\draw   (43.44,45.36) -- (104.24,45.36) -- (104.24,106.16) -- (43.44,106.16) -- cycle ;
%Image [id:dp991104573465732] 
\draw (83.84,255.6) node  {\includegraphics[width=45.6pt,height=45.6pt]{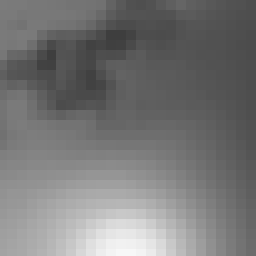}};
%Shape: Rectangle [id:dp9168500499894487] 
\draw   (53.44,225.2) -- (114.24,225.2) -- (114.24,286) -- (53.44,286) -- cycle ;
%Image [id:dp2921782273796041] 
\draw (73.84,265.76) node  {\includegraphics[width=45.6pt,height=45.6pt]{Fig/bunch/3/4_msl_era5_grey.png}};
%Shape: Rectangle [id:dp14977439390228797] 
\draw   (43.44,235.36) -- (104.24,235.36) -- (104.24,296.16) -- (43.44,296.16) -- cycle ;
%Image [id:dp7614995495202921] 
\draw (83.84,159.76) node  {\includegraphics[width=45.6pt,height=45.6pt]{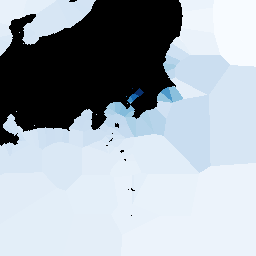}};
%Shape: Rectangle [id:dp04569238864000269] 
\draw   (53.44,129.36) -- (114.24,129.36) -- (114.24,190.16) -- (53.44,190.16) -- cycle ;
%Image [id:dp2794154968385454] 
\draw (73.84,169.76) node  {\includegraphics[width=45.6pt,height=45.6pt]{Fig/bunch/3/gtsm/4_out_model_masked.png}};
%Shape: Rectangle [id:dp5847485781911803] 
\draw   (43.44,139.36) -- (104.24,139.36) -- (104.24,200.16) -- (43.44,200.16) -- cycle ;
%Straight Lines [id:da2796424587449289] 
\draw    (113.04,301.6) -- (140.56,274.99) ;
\draw [shift={(142,273.6)}, rotate = 135.97] [color={rgb, 255:red, 0; green, 0; blue, 0 }  ][line width=0.75]    (10.93,-3.29) .. controls (6.95,-1.4) and (3.31,-0.3) .. (0,0) .. controls (3.31,0.3) and (6.95,1.4) .. (10.93,3.29)   ;
%Shape: Ellipse [id:dp7223231547149207] 
\draw  [fill={rgb, 255:red, 37; green, 37; blue, 37 }  ,fill opacity=0.5 ] (120.4,273.55) .. controls (120.4,272.53) and (121.27,271.7) .. (122.35,271.7) .. controls (123.42,271.7) and (124.3,272.53) .. (124.3,273.55) .. controls (124.3,274.57) and (123.42,275.4) .. (122.35,275.4) .. controls (121.27,275.4) and (120.4,274.57) .. (120.4,273.55) -- cycle ;
%Shape: Ellipse [id:dp8290124492611026] 
\draw  [fill={rgb, 255:red, 37; green, 37; blue, 37 }  ,fill opacity=0.5 ] (124.77,270) .. controls (124.77,268.98) and (125.64,268.15) .. (126.72,268.15) .. controls (127.8,268.15) and (128.67,268.98) .. (128.67,270) .. controls (128.67,271.02) and (127.8,271.84) .. (126.72,271.84) .. controls (125.64,271.84) and (124.77,271.02) .. (124.77,270) -- cycle ;
%Shape: Ellipse [id:dp6183168257340075] 
\draw  [fill={rgb, 255:red, 37; green, 37; blue, 37 }  ,fill opacity=0.5 ] (129.14,265.85) .. controls (129.14,264.83) and (130.02,264) .. (131.09,264) .. controls (132.17,264) and (133.04,264.83) .. (133.04,265.85) .. controls (133.04,266.87) and (132.17,267.7) .. (131.09,267.7) .. controls (130.02,267.7) and (129.14,266.87) .. (129.14,265.85) -- cycle ;
%Shape: Ellipse [id:dp3914168288159803] 
\draw  [fill={rgb, 255:red, 37; green, 37; blue, 37 }  ,fill opacity=0.5 ] (120.4,177.55) .. controls (120.4,176.53) and (121.27,175.7) .. (122.35,175.7) .. controls (123.42,175.7) and (124.3,176.53) .. (124.3,177.55) .. controls (124.3,178.57) and (123.42,179.4) .. (122.35,179.4) .. controls (121.27,179.4) and (120.4,178.57) .. (120.4,177.55) -- cycle ;
%Shape: Ellipse [id:dp060645764745941744] 
\draw  [fill={rgb, 255:red, 37; green, 37; blue, 37 }  ,fill opacity=0.5 ] (124.77,174) .. controls (124.77,172.98) and (125.64,172.15) .. (126.72,172.15) .. controls (127.8,172.15) and (128.67,172.98) .. (128.67,174) .. controls (128.67,175.02) and (127.8,175.84) .. (126.72,175.84) .. controls (125.64,175.84) and (124.77,175.02) .. (124.77,174) -- cycle ;
%Shape: Ellipse [id:dp8395546364347912] 
\draw  [fill={rgb, 255:red, 37; green, 37; blue, 37 }  ,fill opacity=0.5 ] (129.14,169.85) .. controls (129.14,168.83) and (130.02,168) .. (131.09,168) .. controls (132.17,168) and (133.04,168.83) .. (133.04,169.85) .. controls (133.04,170.87) and (132.17,171.7) .. (131.09,171.7) .. controls (130.02,171.7) and (129.14,170.87) .. (129.14,169.85) -- cycle ;
%Shape: Ellipse [id:dp8940361792527816] 
\draw  [fill={rgb, 255:red, 37; green, 37; blue, 37 }  ,fill opacity=0.5 ] (124.77,80) .. controls (124.77,78.98) and (125.64,78.15) .. (126.72,78.15) .. controls (127.8,78.15) and (128.67,78.98) .. (128.67,80) .. controls (128.67,81.02) and (127.8,81.84) .. (126.72,81.84) .. controls (125.64,81.84) and (124.77,81.02) .. (124.77,80) -- cycle ;
%Shape: Ellipse [id:dp7614592229575639] 
\draw  [fill={rgb, 255:red, 37; green, 37; blue, 37 }  ,fill opacity=0.5 ] (129.14,75.85) .. controls (129.14,74.83) and (130.02,74) .. (131.09,74) .. controls (132.17,74) and (133.04,74.83) .. (133.04,75.85) .. controls (133.04,76.87) and (132.17,77.7) .. (131.09,77.7) .. controls (130.02,77.7) and (129.14,76.87) .. (129.14,75.85) -- cycle ;
%Shape: Trapezoid [id:dp5814960922282624] 
\draw   (153.8,34.6) -- (236,66.62) -- (236,260.33) -- (153.8,292.35) -- cycle ;
%Shape: Rectangle [id:dp3907215794083556] 
\draw   (245,170.16) -- (266.67,170.16) -- (266.67,271.6) -- (245,271.6) -- cycle ;
%Shape: Rectangle [id:dp2877959525906373] 
\draw   (299.84,76.96) -- (360.64,76.96) -- (360.64,137.76) -- (299.84,137.76) -- cycle ;
%Shape: Rectangle [id:dp5846049496913888] 
\draw   (299.84,184.96) -- (360.64,184.96) -- (360.64,245.76) -- (299.84,245.76) -- cycle ;
%Shape: Ellipse [id:dp33947761463750714] 
\draw  [fill={rgb, 255:red, 37; green, 37; blue, 37 }  ,fill opacity=0.5 ] (74.4,219.55) .. controls (74.4,218.53) and (75.27,217.7) .. (76.35,217.7) .. controls (77.42,217.7) and (78.3,218.53) .. (78.3,219.55) .. controls (78.3,220.57) and (77.42,221.4) .. (76.35,221.4) .. controls (75.27,221.4) and (74.4,220.57) .. (74.4,219.55) -- cycle ;
%Shape: Ellipse [id:dp6147102621664717] 
\draw  [fill={rgb, 255:red, 37; green, 37; blue, 37 }  ,fill opacity=0.5 ] (74.4,213.55) .. controls (74.4,212.53) and (75.27,211.7) .. (76.35,211.7) .. controls (77.42,211.7) and (78.3,212.53) .. (78.3,213.55) .. controls (78.3,214.57) and (77.42,215.4) .. (76.35,215.4) .. controls (75.27,215.4) and (74.4,214.57) .. (74.4,213.55) -- cycle ;
%Shape: Ellipse [id:dp726103188473175] 
\draw  [fill={rgb, 255:red, 37; green, 37; blue, 37 }  ,fill opacity=0.5 ] (74.4,207.55) .. controls (74.4,206.53) and (75.27,205.7) .. (76.35,205.7) .. controls (77.42,205.7) and (78.3,206.53) .. (78.3,207.55) .. controls (78.3,208.57) and (77.42,209.4) .. (76.35,209.4) .. controls (75.27,209.4) and (74.4,208.57) .. (74.4,207.55) -- cycle ;
%Straight Lines [id:da5974645044936182] 
\draw [line width=1.5]    (121.4,66.8) -- (144.4,66.8) ;
\draw [shift={(147.4,66.8)}, rotate = 180] [color={rgb, 255:red, 0; green, 0; blue, 0 }  ][line width=1.5]    (14.21,-4.28) .. controls (9.04,-1.82) and (4.3,-0.39) .. (0,0) .. controls (4.3,0.39) and (9.04,1.82) .. (14.21,4.28)   ;
%Straight Lines [id:da919971573337679] 
\draw [line width=1.5]    (121.4,161.8) -- (144.4,161.8) ;
\draw [shift={(147.4,161.8)}, rotate = 180] [color={rgb, 255:red, 0; green, 0; blue, 0 }  ][line width=1.5]    (14.21,-4.28) .. controls (9.04,-1.82) and (4.3,-0.39) .. (0,0) .. controls (4.3,0.39) and (9.04,1.82) .. (14.21,4.28)   ;
%Straight Lines [id:da13647720449623568] 
\draw [line width=1.5]    (121.4,253.8) -- (144.4,253.8) ;
\draw [shift={(147.4,253.8)}, rotate = 180] [color={rgb, 255:red, 0; green, 0; blue, 0 }  ][line width=1.5]    (14.21,-4.28) .. controls (9.04,-1.82) and (4.3,-0.39) .. (0,0) .. controls (4.3,0.39) and (9.04,1.82) .. (14.21,4.28)   ;
%Straight Lines [id:da5758057412452724] 
\draw [line width=1.5]    (269.67,215.6) -- (291.4,215.42) ;
\draw [shift={(294.4,215.4)}, rotate = 179.54] [color={rgb, 255:red, 0; green, 0; blue, 0 }  ][line width=1.5]    (14.21,-4.28) .. controls (9.04,-1.82) and (4.3,-0.39) .. (0,0) .. controls (4.3,0.39) and (9.04,1.82) .. (14.21,4.28)   ;
%Shape: Ellipse [id:dp68858258443389] 
\draw  [fill={rgb, 255:red, 37; green, 37; blue, 37 }  ,fill opacity=0.5 ] (119.77,84) .. controls (119.77,82.98) and (120.64,82.15) .. (121.72,82.15) .. controls (122.8,82.15) and (123.67,82.98) .. (123.67,84) .. controls (123.67,85.02) and (122.8,85.84) .. (121.72,85.84) .. controls (120.64,85.84) and (119.77,85.02) .. (119.77,84) -- cycle ;
%Shape: Rectangle [id:dp38506782442079734] 
\draw   (154,307.63) -- (235,307.63) -- (235,348.6) -- (154,348.6) -- cycle ;
%Straight Lines [id:da4774960454272572] 
\draw [line width=1.5]    (195,304.6) -- (195.01,283.1) ;
\draw [shift={(195.01,280.1)}, rotate = 90.02] [color={rgb, 255:red, 0; green, 0; blue, 0 }  ][line width=1.5]    (14.21,-4.28) .. controls (9.04,-1.82) and (4.3,-0.39) .. (0,0) .. controls (4.3,0.39) and (9.04,1.82) .. (14.21,4.28)   ;
%Straight Lines [id:da05142981659188672] 
\draw [line width=1.5]    (121.4,330.8) -- (144.4,330.8) ;
\draw [shift={(147.4,330.8)}, rotate = 180] [color={rgb, 255:red, 0; green, 0; blue, 0 }  ][line width=1.5]    (14.21,-4.28) .. controls (9.04,-1.82) and (4.3,-0.39) .. (0,0) .. controls (4.3,0.39) and (9.04,1.82) .. (14.21,4.28)   ;
%Shape: Rectangle [id:dp8652523579661154] 
\draw   (245,59.16) -- (266.67,59.16) -- (266.67,160.6) -- (245,160.6) -- cycle ;
%Straight Lines [id:da14296005439966253] 
\draw [line width=1.5]    (269,104.6) -- (291.4,104.42) ;
\draw [shift={(294.4,104.4)}, rotate = 179.55] [color={rgb, 255:red, 0; green, 0; blue, 0 }  ][line width=1.5]    (14.21,-4.28) .. controls (9.04,-1.82) and (4.3,-0.39) .. (0,0) .. controls (4.3,0.39) and (9.04,1.82) .. (14.21,4.28)   ;

% Text Node
\draw (168.6,266.92) node [anchor=north west][inner sep=0.75pt]  [font=\Large,rotate=-270] [align=left] {\begin{minipage}[lt]{150.64pt}\setlength\topsep{0pt}
\begin{center}
Spatio-temporal\\deep neural network
\end{center}

\end{minipage}};
% Text Node
\draw (15.4,4.2) node [anchor=north west][inner sep=0.75pt]   [font=\large, align=left] {\textbf{Input} sequence of length $T$};
% Text Node
\draw (245.76,4.28) node [anchor=north west][inner sep=0.75pt]   [font=\large, align=left] {\textbf{Output} for lead time $L$};
% Text Node
\draw (306.64,253.8) node [anchor=north west][inner sep=0.75pt]  [font=\normalsize]  {$(T+L)$};
% Text Node
\draw (3.2,111.92) node [anchor=north west][inner sep=0.75pt]  [font=\normalsize,rotate=-270] [align=left] {\begin{minipage}[lt]{58.61pt}\setlength\topsep{0pt}
\begin{center}
Sparse in-situ\\tide gauges
\end{center}

\end{minipage}};
% Text Node
\draw (3,208.92) node [anchor=north west][inner sep=0.75pt]  [font=\normalsize,rotate=-270] [align=left] {\begin{minipage}[lt]{61.42pt}\setlength\topsep{0pt}
\begin{center}
Coarse ocean\\state reanalysis
\end{center}

\end{minipage}};
% Text Node
\draw (3,304.92) node [anchor=north west][inner sep=0.75pt]  [font=\normalsize,rotate=-270] [align=left] {\begin{minipage}[lt]{61.42pt}\setlength\topsep{0pt}
\begin{center}
Atmospheric\\state reanalysis
\end{center}

\end{minipage}};
% Text Node
\draw (366,152.92) node [anchor=north west][inner sep=0.75pt]  [font=\normalsize,rotate=-270] [align=left] {\begin{minipage}[lt]{65.41pt}\setlength\topsep{0pt}
\begin{center}
Densified storm\\surge forecast
\end{center}

\end{minipage}};
% Text Node
\draw (366,263.92) node [anchor=north west][inner sep=0.75pt]  [font=\normalsize,rotate=-270] [align=left] {\begin{minipage}[lt]{68.32pt}\setlength\topsep{0pt}
\begin{center}
Auxiliary coarse\\state forecast
\end{center}

\end{minipage}};
% Text Node
\draw (110.04,306) node [anchor=north west][inner sep=0.75pt]  [font=\fontsize{0.75em}{0.78em}\selectfont]  {$1$};
% Text Node
\draw (119.84,298) node [anchor=north west][inner sep=0.75pt]  [font=\fontsize{0.75em}{0.78em}\selectfont]  {$2$};
% Text Node
\draw (128.06,297.17) node [anchor=north west][inner sep=0.75pt]  [font=\fontsize{0.75em}{0.78em}\selectfont,rotate=-316.28]  {$...$};
% Text Node
\draw (138.84,281) node [anchor=north west][inner sep=0.75pt]  [font=\fontsize{0.75em}{0.78em}\selectfont]  {$T$};
% Text Node
\draw (249,265.32) node [anchor=north west][inner sep=0.75pt]  [font=\fontsize{1.05em}{1.12em}\selectfont,rotate=-270] [align=left] {\begin{minipage}[lt]{64.09pt}\setlength\topsep{0pt}
\begin{center}
weight sharing
\end{center}

\end{minipage}};
% Text Node
\draw (156,313.63) node [anchor=north west][inner sep=0.75pt]  [font=\fontsize{1.1em}{1.12em}\selectfont] [align=left] {\begin{minipage}[lt]{53.52pt}\setlength\topsep{0pt}
\begin{center}
temporal\\conditioning
\end{center}

\end{minipage}};
% Text Node
\draw (36.76,323) node [anchor=north west][inner sep=0.75pt]   [font=\fontsize{1.05em}{1.12em}, align=left] {lead time $L$};
% Text Node
\draw (249,156.32) node [anchor=north west][inner sep=0.75pt]  [font=\fontsize{1.05em}{1.12em}\selectfont,rotate=-270] [align=left] {\begin{minipage}[lt]{64.09pt}\setlength\topsep{0pt}
\begin{center}
weight sharing
\end{center}

\end{minipage}};

\end{tikzpicture}

%% file: Fig/exp_setup.tex
\tikzset{every picture/.style={line width=0.75pt}} %set default line width to 0.75pt        

\begin{tikzpicture}[x=0.75pt,y=0.75pt,yscale=-1,xscale=1]
%uncomment if require: \path (0,381); %set diagram left start at 0, and has height of 381

%Shape: Polygon Curved [id:ds04042706462992718] 
\draw  [color={rgb, 255:red, 37; green, 37; blue, 37 }  ,draw opacity=1 ][fill={rgb, 255:red, 196; green, 164; blue, 132 }  ,fill opacity=0.29 ] (46.34,86.52) .. controls (47.34,84.9) and (60.65,79.72) .. (62.32,79.72) .. controls (63.98,79.72) and (66.6,87.61) .. (66.98,89.44) .. controls (67.35,91.27) and (70.64,116.65) .. (68.97,118.59) .. controls (67.31,120.53) and (62.32,117.94) .. (61.65,119.23) .. controls (60.99,120.53) and (53.33,126.68) .. (52.67,125.06) .. controls (52,123.44) and (50.6,121.11) .. (49.67,116.64) .. controls (48.74,112.18) and (48.5,105.58) .. (48.67,102.72) .. controls (48.85,99.85) and (45.34,88.14) .. (46.34,86.52) -- cycle ;
%Shape: Polygon Curved [id:ds7703353776618984] 
\draw  [color={rgb, 255:red, 37; green, 37; blue, 37 }  ,draw opacity=1 ][fill={rgb, 255:red, 196; green, 164; blue, 132 }  ,fill opacity=0.29 ] (46.68,50.9) .. controls (48.11,49.4) and (49,47.33) .. (52,47.66) .. controls (55,47.98) and (54.57,48.33) .. (54.88,47.27) .. controls (55.19,46.21) and (54.64,43.29) .. (55,43.12) .. controls (55.35,42.95) and (57.32,47.98) .. (57.32,51.87) .. controls (57.32,55.75) and (58.66,55.75) .. (58.32,58.67) .. controls (57.99,61.58) and (55.33,68.71) .. (53.66,68.39) .. controls (52,68.06) and (48.55,65.56) .. (46.68,62.23) .. controls (44.8,58.9) and (45.24,52.4) .. (46.68,50.9) -- cycle ;
%Curve Lines [id:da13021754390563522] 
\draw [color={rgb, 255:red, 37; green, 37; blue, 37 }  ,draw opacity=1 ][fill={rgb, 255:red, 196; green, 164; blue, 132 }  ,fill opacity=0.29 ]   (133.77,40.13) .. controls (119.43,36.8) and (95.43,42.47) .. (95.43,48.8) .. controls (95.43,55.13) and (113.16,68.25) .. (120.55,72.57) .. controls (127.94,76.88) and (128.67,104.11) .. (132,103.46) .. controls (135.33,102.81) and (156.82,90.06) .. (161.96,92.86) .. controls (167.11,95.65) and (159.08,119.29) .. (159.1,122.07) .. controls (159.12,124.84) and (183.74,123.21) .. (182.78,128.92) .. controls (181.82,134.63) and (179.06,140.22) .. (181.45,140.9) .. controls (183.84,141.58) and (202.75,122.44) .. (199.42,113.37) .. controls (196.09,104.31) and (217.06,105.28) .. (218.39,101.72) .. controls (219.72,98.15) and (187.69,93.07) .. (188.1,90.47) .. controls (188.51,87.86) and (212.22,84.32) .. (201.42,79.04) .. controls (190.62,73.77) and (181.45,74.19) .. (178.45,69.98) .. controls (175.46,65.77) and (179.99,65.29) .. (177.46,62.85) .. controls (174.92,60.42) and (156.05,58.27) .. (150.17,53.78) .. controls (144.28,49.29) and (140.57,50.13) .. (133.77,40.13) -- cycle ;
%Shape: Polygon Curved [id:ds058754076911435416] 
\draw  [color={rgb, 255:red, 37; green, 37; blue, 37 }  ,draw opacity=1 ][fill={rgb, 255:red, 196; green, 164; blue, 132 }  ,fill opacity=0.29 ] (122.55,144.17) .. controls (124.52,143.02) and (171.8,137.37) .. (171.8,139.64) .. controls (171.8,141.9) and (169.64,146.15) .. (169.14,151.94) .. controls (168.63,157.74) and (169.42,166.63) .. (168.14,167.16) .. controls (166.95,167.66) and (162.96,165.98) .. (158.23,163.78) .. controls (157.87,163.61) and (157.52,163.45) .. (157.16,163.28) .. controls (151.96,160.84) and (149.25,158.96) .. (146.18,157.77) .. controls (143.11,156.59) and (136.24,156.17) .. (131.53,153.89) .. controls (126.82,151.6) and (120.58,145.32) .. (122.55,144.17) -- cycle ;
%Shape: Square [id:dp8619928343375745] 
\draw  [fill={rgb, 255:red, 195; green, 218; blue, 255 }  ,fill opacity=0.6 ] (94.4,52.6) -- (103.4,52.6) -- (103.4,61.6) -- (94.4,61.6) -- cycle ;
%Shape: Square [id:dp7805393993415093] 
\draw  [fill={rgb, 255:red, 195; green, 218; blue, 255 }  ,fill opacity=0.6 ] (116.88,73.57) -- (125.88,73.57) -- (125.88,82.57) -- (116.88,82.57) -- cycle ;
%Shape: Square [id:dp06901348140708641] 
\draw  [fill={rgb, 255:red, 195; green, 218; blue, 255 }  ,fill opacity=0.6 ] (154.63,88.86) -- (163.63,88.86) -- (163.63,97.86) -- (154.63,97.86) -- cycle ;
%Shape: Square [id:dp03485272759235536] 
\draw  [fill={rgb, 255:red, 195; green, 218; blue, 255 }  ,fill opacity=0.6 ] (156.63,115.86) -- (165.63,115.86) -- (165.63,124.86) -- (156.63,124.86) -- cycle ;
%Shape: Square [id:dp39080017679535173] 
\draw  [fill={rgb, 255:red, 195; green, 218; blue, 255 }  ,fill opacity=0.6 ] (162.8,139.64) -- (171.8,139.64) -- (171.8,148.64) -- (162.8,148.64) -- cycle ;
%Shape: Square [id:dp2963943663195847] 
\draw  [fill={rgb, 255:red, 195; green, 218; blue, 255 }  ,fill opacity=0.6 ] (52.65,119.23) -- (61.65,119.23) -- (61.65,128.23) -- (52.65,128.23) -- cycle ;
%Shape: Square [id:dp7251365884610583] 
\draw  [fill={rgb, 255:red, 195; green, 218; blue, 255 }  ,fill opacity=0.6 ] (64.65,107.23) -- (73.65,107.23) -- (73.65,116.23) -- (64.65,116.23) -- cycle ;
%Shape: Square [id:dp13132534356007697] 
\draw  [fill={rgb, 255:red, 195; green, 218; blue, 255 }  ,fill opacity=0.6 ] (44.17,98.22) -- (53.17,98.22) -- (53.17,107.22) -- (44.17,107.22) -- cycle ;
%Shape: Square [id:dp23178310851366635] 
\draw  [fill={rgb, 255:red, 195; green, 218; blue, 255 }  ,fill opacity=0.6 ] (122.53,144.89) -- (131.53,144.89) -- (131.53,153.89) -- (122.53,153.89) -- cycle ;
%Shape: Square [id:dp6205038447259348] 
\draw  [fill={rgb, 255:red, 195; green, 218; blue, 255 }  ,fill opacity=0.6 ] (214.89,98.22) -- (223.89,98.22) -- (223.89,107.22) -- (214.89,107.22) -- cycle ;
%Shape: Square [id:dp9535062693027643] 
\draw  [fill={rgb, 255:red, 195; green, 218; blue, 255 }  ,fill opacity=0.6 ] (192.49,125.42) -- (201.49,125.42) -- (201.49,134.42) -- (192.49,134.42) -- cycle ;
%Shape: Square [id:dp2662557581948637] 
\draw  [fill={rgb, 255:red, 195; green, 218; blue, 255 }  ,fill opacity=0.6 ] (185.29,69.42) -- (194.29,69.42) -- (194.29,78.42) -- (185.29,78.42) -- cycle ;
%Shape: Square [id:dp11230035710881947] 
\draw  [fill={rgb, 255:red, 195; green, 218; blue, 255 }  ,fill opacity=0.6 ] (42.09,51.82) -- (51.09,51.82) -- (51.09,60.82) -- (42.09,60.82) -- cycle ;
%Shape: Square [id:dp10665729821478309] 
\draw  [fill={rgb, 255:red, 195; green, 218; blue, 255 }  ,fill opacity=0.6 ] (53.66,59.39) -- (62.66,59.39) -- (62.66,68.39) -- (53.66,68.39) -- cycle ;
%Shape: Square [id:dp27157643551681065] 
\draw  [fill={rgb, 255:red, 195; green, 218; blue, 255 }  ,fill opacity=0.6 ] (142.17,45.78) -- (151.17,45.78) -- (151.17,54.78) -- (142.17,54.78) -- cycle ;
%Shape: Square [id:dp1404561387592873] 
\draw  [fill={rgb, 255:red, 195; green, 218; blue, 255 }  ,fill opacity=0.6 ] (163.77,56.18) -- (172.77,56.18) -- (172.77,65.18) -- (163.77,65.18) -- cycle ;
%Shape: Circle [id:dp3533886950562759] 
\draw  [fill={rgb, 255:red, 217; green, 56; blue, 127 }  ,fill opacity=0.6 ] (54.08,46.47) .. controls (54.08,43.87) and (56.18,41.77) .. (58.78,41.77) .. controls (61.37,41.77) and (63.48,43.87) .. (63.48,46.47) .. controls (63.48,49.07) and (61.37,51.17) .. (58.78,51.17) .. controls (56.18,51.17) and (54.08,49.07) .. (54.08,46.47) -- cycle ;
%Shape: Circle [id:dp26196732864484407] 
\draw  [fill={rgb, 255:red, 217; green, 56; blue, 127 }  ,fill opacity=0.6 ] (61.48,93.34) .. controls (61.48,90.74) and (63.58,88.64) .. (66.18,88.64) .. controls (68.77,88.64) and (70.88,90.74) .. (70.88,93.34) .. controls (70.88,95.93) and (68.77,98.04) .. (66.18,98.04) .. controls (63.58,98.04) and (61.48,95.93) .. (61.48,93.34) -- cycle ;
%Shape: Circle [id:dp6293530131127898] 
\draw  [fill={rgb, 255:red, 217; green, 56; blue, 127 }  ,fill opacity=0.6 ] (156.36,162.48) .. controls (156.36,159.88) and (158.46,157.78) .. (161.06,157.78) .. controls (163.65,157.78) and (165.76,159.88) .. (165.76,162.48) .. controls (165.76,165.07) and (163.65,167.18) .. (161.06,167.18) .. controls (158.46,167.18) and (156.36,165.07) .. (156.36,162.48) -- cycle ;
%Shape: Circle [id:dp1038441623549875] 
\draw  [fill={rgb, 255:red, 217; green, 56; blue, 127 }  ,fill opacity=0.6 ] (190.89,86.34) .. controls (190.89,83.75) and (193,81.64) .. (195.59,81.64) .. controls (198.19,81.64) and (200.29,83.75) .. (200.29,86.34) .. controls (200.29,88.94) and (198.19,91.04) .. (195.59,91.04) .. controls (193,91.04) and (190.89,88.94) .. (190.89,86.34) -- cycle ;
%Shape: Circle [id:dp010134306301013307] 
\draw  [fill={rgb, 255:red, 217; green, 56; blue, 127 }  ,fill opacity=0.6 ] (101.96,64.08) .. controls (101.96,61.48) and (104.06,59.38) .. (106.66,59.38) .. controls (109.25,59.38) and (111.36,61.48) .. (111.36,64.08) .. controls (111.36,66.67) and (109.25,68.78) .. (106.66,68.78) .. controls (104.06,68.78) and (101.96,66.67) .. (101.96,64.08) -- cycle ;
%Shape: Circle [id:dp5817304313567007] 
\draw  [fill={rgb, 255:red, 217; green, 56; blue, 127 }  ,fill opacity=0.6 ] (127.3,103.46) .. controls (127.3,100.86) and (129.4,98.76) .. (132,98.76) .. controls (134.6,98.76) and (136.7,100.86) .. (136.7,103.46) .. controls (136.7,106.06) and (134.6,108.16) .. (132,108.16) .. controls (129.4,108.16) and (127.3,106.06) .. (127.3,103.46) -- cycle ;
%Shape: Circle [id:dp9486370555492536] 
\draw  [fill={rgb, 255:red, 217; green, 56; blue, 127 }  ,fill opacity=0.6 ] (178.28,133.82) .. controls (178.28,131.22) and (180.38,129.12) .. (182.98,129.12) .. controls (185.58,129.12) and (187.68,131.22) .. (187.68,133.82) .. controls (187.68,136.42) and (185.58,138.52) .. (182.98,138.52) .. controls (180.38,138.52) and (178.28,136.42) .. (178.28,133.82) -- cycle ;
%Shape: Circle [id:dp863773075437404] 
\draw  [fill={rgb, 255:red, 217; green, 56; blue, 127 }  ,fill opacity=0.6 ] (107.41,42.09) .. controls (107.41,39.49) and (109.52,37.39) .. (112.11,37.39) .. controls (114.71,37.39) and (116.81,39.49) .. (116.81,42.09) .. controls (116.81,44.68) and (114.71,46.79) .. (112.11,46.79) .. controls (109.52,46.79) and (107.41,44.68) .. (107.41,42.09) -- cycle ;
%Shape: Circle [id:dp1544187108487476] 
\draw  [fill={rgb, 255:red, 217; green, 56; blue, 127 }  ,fill opacity=0.6 ] (153.48,57.82) .. controls (153.48,55.22) and (155.58,53.12) .. (158.18,53.12) .. controls (160.78,53.12) and (162.88,55.22) .. (162.88,57.82) .. controls (162.88,60.42) and (160.78,62.52) .. (158.18,62.52) .. controls (155.58,62.52) and (153.48,60.42) .. (153.48,57.82) -- cycle ;
%Shape: Rectangle [id:dp29224558941337864] 
\draw   (32,31.4) -- (232,31.4) -- (232,175.52) -- (32,175.52) -- cycle ;
%Shape: Circle [id:dp5080777328806794] 
\draw  [fill={rgb, 255:red, 217; green, 56; blue, 127 }  ,fill opacity=0.6 ] (277.6,294.26) .. controls (277.6,291.66) and (279.7,289.56) .. (282.3,289.56) .. controls (284.9,289.56) and (287,291.66) .. (287,294.26) .. controls (287,296.85) and (284.9,298.96) .. (282.3,298.96) .. controls (279.7,298.96) and (277.6,296.85) .. (277.6,294.26) -- cycle ;
%Shape: Rectangle [id:dp4695821671154621] 
\draw   (272.4,261.4) -- (396.73,261.4) -- (396.73,305.8) -- (272.4,305.8) -- cycle ;
%Shape: Square [id:dp6913346054545051] 
\draw  [fill={rgb, 255:red, 195; green, 218; blue, 255 }  ,fill opacity=0.6 ] (277.69,267.82) -- (286.69,267.82) -- (286.69,276.82) -- (277.69,276.82) -- cycle ;
%Shape: Polygon Curved [id:ds555789808121939] 
\draw  [color={rgb, 255:red, 37; green, 37; blue, 37 }  ,draw opacity=1 ][fill={rgb, 255:red, 196; green, 164; blue, 132 }  ,fill opacity=0.29 ] (266.34,156.52) .. controls (267.34,154.9) and (280.65,149.72) .. (282.32,149.72) .. controls (283.98,149.72) and (286.6,157.61) .. (286.98,159.44) .. controls (287.35,161.27) and (290.64,186.65) .. (288.97,188.59) .. controls (287.31,190.53) and (282.32,187.94) .. (281.65,189.23) .. controls (280.99,190.53) and (273.33,196.68) .. (272.67,195.06) .. controls (272,193.44) and (270.6,191.11) .. (269.67,186.64) .. controls (268.74,182.18) and (268.5,175.58) .. (268.67,172.72) .. controls (268.85,169.85) and (265.34,158.14) .. (266.34,156.52) -- cycle ;
%Shape: Polygon Curved [id:ds05208103987001955] 
\draw  [color={rgb, 255:red, 37; green, 37; blue, 37 }  ,draw opacity=1 ][fill={rgb, 255:red, 196; green, 164; blue, 132 }  ,fill opacity=0.29 ] (266.68,120.9) .. controls (268.11,119.4) and (269,117.33) .. (272,117.66) .. controls (275,117.98) and (274.57,118.33) .. (274.88,117.27) .. controls (275.19,116.21) and (274.64,113.29) .. (275,113.12) .. controls (275.35,112.95) and (277.32,117.98) .. (277.32,121.87) .. controls (277.32,125.75) and (278.66,125.75) .. (278.32,128.67) .. controls (277.99,131.58) and (275.33,138.71) .. (273.66,138.39) .. controls (272,138.06) and (268.55,135.56) .. (266.68,132.23) .. controls (264.8,128.9) and (265.24,122.4) .. (266.68,120.9) -- cycle ;
%Curve Lines [id:da2822239152783468] 
\draw [color={rgb, 255:red, 37; green, 37; blue, 37 }  ,draw opacity=1 ][fill={rgb, 255:red, 196; green, 164; blue, 132 }  ,fill opacity=0.29 ]   (353.77,110.13) .. controls (339.43,106.8) and (315.43,112.47) .. (315.43,118.8) .. controls (315.43,125.13) and (333.16,138.25) .. (340.55,142.57) .. controls (347.94,146.88) and (348.67,174.11) .. (352,173.46) .. controls (355.33,172.81) and (376.82,160.06) .. (381.96,162.86) .. controls (387.11,165.65) and (379.08,189.29) .. (379.1,192.07) .. controls (379.12,194.84) and (403.74,193.21) .. (402.78,198.92) .. controls (401.82,204.63) and (399.06,210.22) .. (401.45,210.9) .. controls (403.84,211.58) and (422.75,192.44) .. (419.42,183.37) .. controls (416.09,174.31) and (437.06,175.28) .. (438.39,171.72) .. controls (439.72,168.15) and (407.69,163.07) .. (408.1,160.47) .. controls (408.51,157.86) and (432.22,154.32) .. (421.42,149.04) .. controls (410.62,143.77) and (401.45,144.19) .. (398.45,139.98) .. controls (395.46,135.77) and (399.99,135.29) .. (397.46,132.85) .. controls (394.92,130.42) and (376.05,128.27) .. (370.17,123.78) .. controls (364.28,119.29) and (360.57,120.13) .. (353.77,110.13) -- cycle ;
%Shape: Polygon Curved [id:ds3916739379899472] 
\draw  [color={rgb, 255:red, 37; green, 37; blue, 37 }  ,draw opacity=1 ][fill={rgb, 255:red, 196; green, 164; blue, 132 }  ,fill opacity=0.29 ] (342.55,214.17) .. controls (344.52,213.02) and (391.8,207.37) .. (391.8,209.64) .. controls (391.8,211.9) and (389.64,216.15) .. (389.14,221.94) .. controls (388.63,227.74) and (389.42,236.63) .. (388.14,237.16) .. controls (386.95,237.66) and (382.96,235.98) .. (378.23,233.78) .. controls (377.87,233.61) and (377.52,233.45) .. (377.16,233.28) .. controls (371.96,230.84) and (369.25,228.96) .. (366.18,227.77) .. controls (363.11,226.59) and (356.24,226.17) .. (351.53,223.89) .. controls (346.82,221.6) and (340.58,215.32) .. (342.55,214.17) -- cycle ;
%Shape: Circle [id:dp38829507745661274] 
\draw  [fill={rgb, 255:red, 217; green, 56; blue, 127 }  ,fill opacity=0.6 ] (274.08,116.47) .. controls (274.08,113.87) and (276.18,111.77) .. (278.78,111.77) .. controls (281.37,111.77) and (283.48,113.87) .. (283.48,116.47) .. controls (283.48,119.07) and (281.37,121.17) .. (278.78,121.17) .. controls (276.18,121.17) and (274.08,119.07) .. (274.08,116.47) -- cycle ;
%Shape: Circle [id:dp8378097203498116] 
\draw  [fill={rgb, 255:red, 217; green, 56; blue, 127 }  ,fill opacity=0.6 ] (281.48,163.34) .. controls (281.48,160.74) and (283.58,158.64) .. (286.18,158.64) .. controls (288.77,158.64) and (290.88,160.74) .. (290.88,163.34) .. controls (290.88,165.93) and (288.77,168.04) .. (286.18,168.04) .. controls (283.58,168.04) and (281.48,165.93) .. (281.48,163.34) -- cycle ;
%Shape: Circle [id:dp7182465804556022] 
\draw  [fill={rgb, 255:red, 217; green, 56; blue, 127 }  ,fill opacity=0.6 ] (376.36,232.48) .. controls (376.36,229.88) and (378.46,227.78) .. (381.06,227.78) .. controls (383.65,227.78) and (385.76,229.88) .. (385.76,232.48) .. controls (385.76,235.07) and (383.65,237.18) .. (381.06,237.18) .. controls (378.46,237.18) and (376.36,235.07) .. (376.36,232.48) -- cycle ;
%Shape: Circle [id:dp7400686577032323] 
\draw  [fill={rgb, 255:red, 217; green, 56; blue, 127 }  ,fill opacity=0.6 ] (410.89,156.34) .. controls (410.89,153.75) and (413,151.64) .. (415.59,151.64) .. controls (418.19,151.64) and (420.29,153.75) .. (420.29,156.34) .. controls (420.29,158.94) and (418.19,161.04) .. (415.59,161.04) .. controls (413,161.04) and (410.89,158.94) .. (410.89,156.34) -- cycle ;
%Shape: Circle [id:dp31597082050815883] 
\draw  [fill={rgb, 255:red, 217; green, 56; blue, 127 }  ,fill opacity=0.6 ] (321.96,134.08) .. controls (321.96,131.48) and (324.06,129.38) .. (326.66,129.38) .. controls (329.25,129.38) and (331.36,131.48) .. (331.36,134.08) .. controls (331.36,136.67) and (329.25,138.78) .. (326.66,138.78) .. controls (324.06,138.78) and (321.96,136.67) .. (321.96,134.08) -- cycle ;
%Shape: Circle [id:dp9020314270122451] 
\draw  [fill={rgb, 255:red, 217; green, 56; blue, 127 }  ,fill opacity=0.6 ] (347.3,173.46) .. controls (347.3,170.86) and (349.4,168.76) .. (352,168.76) .. controls (354.6,168.76) and (356.7,170.86) .. (356.7,173.46) .. controls (356.7,176.06) and (354.6,178.16) .. (352,178.16) .. controls (349.4,178.16) and (347.3,176.06) .. (347.3,173.46) -- cycle ;
%Shape: Circle [id:dp8264166589809698] 
\draw  [fill={rgb, 255:red, 217; green, 56; blue, 127 }  ,fill opacity=0.6 ] (398.28,203.82) .. controls (398.28,201.22) and (400.38,199.12) .. (402.98,199.12) .. controls (405.58,199.12) and (407.68,201.22) .. (407.68,203.82) .. controls (407.68,206.42) and (405.58,208.52) .. (402.98,208.52) .. controls (400.38,208.52) and (398.28,206.42) .. (398.28,203.82) -- cycle ;
%Shape: Circle [id:dp40559438740923026] 
\draw  [fill={rgb, 255:red, 217; green, 56; blue, 127 }  ,fill opacity=0.6 ] (327.41,112.09) .. controls (327.41,109.49) and (329.52,107.39) .. (332.11,107.39) .. controls (334.71,107.39) and (336.81,109.49) .. (336.81,112.09) .. controls (336.81,114.68) and (334.71,116.79) .. (332.11,116.79) .. controls (329.52,116.79) and (327.41,114.68) .. (327.41,112.09) -- cycle ;
%Shape: Circle [id:dp3776714601159197] 
\draw  [fill={rgb, 255:red, 217; green, 56; blue, 127 }  ,fill opacity=0.6 ] (373.48,127.82) .. controls (373.48,125.22) and (375.58,123.12) .. (378.18,123.12) .. controls (380.78,123.12) and (382.88,125.22) .. (382.88,127.82) .. controls (382.88,130.42) and (380.78,132.52) .. (378.18,132.52) .. controls (375.58,132.52) and (373.48,130.42) .. (373.48,127.82) -- cycle ;
%Shape: Rectangle [id:dp852963192915603] 
\draw   (252,101.4) -- (452,101.4) -- (452,245.52) -- (252,245.52) -- cycle ;
%Shape: Polygon Curved [id:ds38700193430773266] 
\draw  [color={rgb, 255:red, 37; green, 37; blue, 37 }  ,draw opacity=1 ][fill={rgb, 255:red, 196; green, 164; blue, 132 }  ,fill opacity=0.29 ] (46.34,246.52) .. controls (47.34,244.9) and (60.65,239.72) .. (62.32,239.72) .. controls (63.98,239.72) and (66.6,247.61) .. (66.98,249.44) .. controls (67.35,251.27) and (70.64,276.65) .. (68.97,278.59) .. controls (67.31,280.53) and (62.32,277.94) .. (61.65,279.23) .. controls (60.99,280.53) and (53.33,286.68) .. (52.67,285.06) .. controls (52,283.44) and (50.6,281.11) .. (49.67,276.64) .. controls (48.74,272.18) and (48.5,265.58) .. (48.67,262.72) .. controls (48.85,259.85) and (45.34,248.14) .. (46.34,246.52) -- cycle ;
%Shape: Polygon Curved [id:ds488434963536724] 
\draw  [color={rgb, 255:red, 37; green, 37; blue, 37 }  ,draw opacity=1 ][fill={rgb, 255:red, 196; green, 164; blue, 132 }  ,fill opacity=0.29 ] (46.68,210.9) .. controls (48.11,209.4) and (49,207.33) .. (52,207.66) .. controls (55,207.98) and (54.57,208.33) .. (54.88,207.27) .. controls (55.19,206.21) and (54.64,203.29) .. (55,203.12) .. controls (55.35,202.95) and (57.32,207.98) .. (57.32,211.87) .. controls (57.32,215.75) and (58.66,215.75) .. (58.32,218.67) .. controls (57.99,221.58) and (55.33,228.71) .. (53.66,228.39) .. controls (52,228.06) and (48.55,225.56) .. (46.68,222.23) .. controls (44.8,218.9) and (45.24,212.4) .. (46.68,210.9) -- cycle ;
%Curve Lines [id:da5473252022731276] 
\draw [color={rgb, 255:red, 37; green, 37; blue, 37 }  ,draw opacity=1 ][fill={rgb, 255:red, 196; green, 164; blue, 132 }  ,fill opacity=0.29 ]   (133.77,200.13) .. controls (119.43,196.8) and (95.43,202.47) .. (95.43,208.8) .. controls (95.43,215.13) and (113.16,228.25) .. (120.55,232.57) .. controls (127.94,236.88) and (128.67,264.11) .. (132,263.46) .. controls (135.33,262.81) and (156.82,250.06) .. (161.96,252.86) .. controls (167.11,255.65) and (159.08,279.29) .. (159.1,282.07) .. controls (159.12,284.84) and (183.74,283.21) .. (182.78,288.92) .. controls (181.82,294.63) and (179.06,300.22) .. (181.45,300.9) .. controls (183.84,301.58) and (202.75,282.44) .. (199.42,273.37) .. controls (196.09,264.31) and (217.06,265.28) .. (218.39,261.72) .. controls (219.72,258.15) and (187.69,253.07) .. (188.1,250.47) .. controls (188.51,247.86) and (212.22,244.32) .. (201.42,239.04) .. controls (190.62,233.77) and (181.45,234.19) .. (178.45,229.98) .. controls (175.46,225.77) and (179.99,225.29) .. (177.46,222.85) .. controls (174.92,220.42) and (156.05,218.27) .. (150.17,213.78) .. controls (144.28,209.29) and (140.57,210.13) .. (133.77,200.13) -- cycle ;
%Shape: Polygon Curved [id:ds09326838281325411] 
\draw  [color={rgb, 255:red, 37; green, 37; blue, 37 }  ,draw opacity=1 ][fill={rgb, 255:red, 196; green, 164; blue, 132 }  ,fill opacity=0.29 ] (122.55,304.17) .. controls (124.52,303.02) and (171.8,297.37) .. (171.8,299.64) .. controls (171.8,301.9) and (169.64,306.15) .. (169.14,311.94) .. controls (168.63,317.74) and (169.42,326.63) .. (168.14,327.16) .. controls (166.95,327.66) and (162.96,325.98) .. (158.23,323.78) .. controls (157.87,323.61) and (157.52,323.45) .. (157.16,323.28) .. controls (151.96,320.84) and (149.25,318.96) .. (146.18,317.77) .. controls (143.11,316.59) and (136.24,316.17) .. (131.53,313.89) .. controls (126.82,311.6) and (120.58,305.32) .. (122.55,304.17) -- cycle ;
%Shape: Square [id:dp036899396479598945] 
\draw  [fill={rgb, 255:red, 195; green, 218; blue, 255 }  ,fill opacity=0.6 ] (94.4,212.6) -- (103.4,212.6) -- (103.4,221.6) -- (94.4,221.6) -- cycle ;
%Shape: Square [id:dp1538644106803262] 
\draw  [fill={rgb, 255:red, 195; green, 218; blue, 255 }  ,fill opacity=0.6 ] (116.88,233.57) -- (125.88,233.57) -- (125.88,242.57) -- (116.88,242.57) -- cycle ;
%Shape: Square [id:dp5533262061495283] 
\draw  [fill={rgb, 255:red, 195; green, 218; blue, 255 }  ,fill opacity=0.6 ] (154.63,248.86) -- (163.63,248.86) -- (163.63,257.86) -- (154.63,257.86) -- cycle ;
%Shape: Square [id:dp7955178475862872] 
\draw  [fill={rgb, 255:red, 195; green, 218; blue, 255 }  ,fill opacity=0.6 ] (156.63,275.86) -- (165.63,275.86) -- (165.63,284.86) -- (156.63,284.86) -- cycle ;
%Shape: Square [id:dp8291450227902541] 
\draw  [fill={rgb, 255:red, 195; green, 218; blue, 255 }  ,fill opacity=0.6 ] (162.8,299.64) -- (171.8,299.64) -- (171.8,308.64) -- (162.8,308.64) -- cycle ;
%Shape: Square [id:dp2221839609643752] 
\draw  [fill={rgb, 255:red, 195; green, 218; blue, 255 }  ,fill opacity=0.6 ] (52.65,279.23) -- (61.65,279.23) -- (61.65,288.23) -- (52.65,288.23) -- cycle ;
%Shape: Square [id:dp8767303288091537] 
\draw  [fill={rgb, 255:red, 195; green, 218; blue, 255 }  ,fill opacity=0.6 ] (64.65,267.23) -- (73.65,267.23) -- (73.65,276.23) -- (64.65,276.23) -- cycle ;
%Shape: Square [id:dp98835291228298] 
\draw  [fill={rgb, 255:red, 195; green, 218; blue, 255 }  ,fill opacity=0.6 ] (44.17,258.22) -- (53.17,258.22) -- (53.17,267.22) -- (44.17,267.22) -- cycle ;
%Shape: Square [id:dp08763552229097193] 
\draw  [fill={rgb, 255:red, 195; green, 218; blue, 255 }  ,fill opacity=0.6 ] (122.53,304.89) -- (131.53,304.89) -- (131.53,313.89) -- (122.53,313.89) -- cycle ;
%Shape: Square [id:dp3843113822107751] 
\draw  [fill={rgb, 255:red, 195; green, 218; blue, 255 }  ,fill opacity=0.6 ] (214.89,258.22) -- (223.89,258.22) -- (223.89,267.22) -- (214.89,267.22) -- cycle ;
%Shape: Square [id:dp07928363991714948] 
\draw  [fill={rgb, 255:red, 195; green, 218; blue, 255 }  ,fill opacity=0.6 ] (192.49,285.42) -- (201.49,285.42) -- (201.49,294.42) -- (192.49,294.42) -- cycle ;
%Shape: Square [id:dp02358155703517273] 
\draw  [fill={rgb, 255:red, 195; green, 218; blue, 255 }  ,fill opacity=0.6 ] (185.29,229.42) -- (194.29,229.42) -- (194.29,238.42) -- (185.29,238.42) -- cycle ;
%Shape: Square [id:dp3755989607413417] 
\draw  [fill={rgb, 255:red, 195; green, 218; blue, 255 }  ,fill opacity=0.6 ] (42.09,211.82) -- (51.09,211.82) -- (51.09,220.82) -- (42.09,220.82) -- cycle ;
%Shape: Square [id:dp4922084920970293] 
\draw  [fill={rgb, 255:red, 195; green, 218; blue, 255 }  ,fill opacity=0.6 ] (53.66,219.39) -- (62.66,219.39) -- (62.66,228.39) -- (53.66,228.39) -- cycle ;
%Shape: Square [id:dp6208496693692414] 
\draw  [fill={rgb, 255:red, 195; green, 218; blue, 255 }  ,fill opacity=0.6 ] (142.17,205.78) -- (151.17,205.78) -- (151.17,214.78) -- (142.17,214.78) -- cycle ;
%Shape: Square [id:dp43145794352510647] 
\draw  [fill={rgb, 255:red, 195; green, 218; blue, 255 }  ,fill opacity=0.6 ] (163.77,216.18) -- (172.77,216.18) -- (172.77,225.18) -- (163.77,225.18) -- cycle ;
%Shape: Rectangle [id:dp06925210352241251] 
\draw   (32,191.4) -- (232,191.4) -- (232,335.52) -- (32,335.52) -- cycle ;
%Curve Lines [id:da3306878294212463] 
\draw [color={rgb, 255:red, 37; green, 37; blue, 37 }  ,draw opacity=0.5 ][line width=1.5]    (98.07,283.38) .. controls (97.94,284.09) and (93.37,279.18) .. (96.9,275.24) .. controls (100.43,271.3) and (117.77,273.14) .. (115.12,282.59) .. controls (112.48,292.04) and (95.65,293.28) .. (88.96,287.31) .. controls (82.28,281.34) and (85.06,272.21) .. (88.56,269.39) .. controls (92.07,266.56) and (106.85,261.86) .. (122.35,267.26) ;
%Shape: Boxed Bezier Curve [id:dp4655899672065422] 
\draw [color={rgb, 255:red, 37; green, 37; blue, 37 }  ,draw opacity=0.5 ][line width=1.5]    (107.7,277.61) .. controls (107.83,276.89) and (112.4,281.81) .. (108.87,285.74) .. controls (105.35,289.68) and (88.01,287.84) .. (90.65,278.39) .. controls (93.3,268.94) and (110.13,267.7) .. (116.81,273.67) .. controls (123.49,279.64) and (120.72,288.78) .. (117.21,291.6) .. controls (113.71,294.42) and (98.92,299.13) .. (83.42,293.72) ;

%Curve Lines [id:da05219821817574943] 
\draw [color={rgb, 255:red, 37; green, 37; blue, 37 }  ,draw opacity=0.5 ][line width=1.5]    (97.07,122.38) .. controls (96.94,123.09) and (92.37,118.18) .. (95.9,114.24) .. controls (99.43,110.3) and (116.77,112.14) .. (114.12,121.59) .. controls (111.48,131.04) and (94.65,132.28) .. (87.96,126.31) .. controls (81.28,120.34) and (84.06,111.21) .. (87.56,108.39) .. controls (91.07,105.56) and (105.85,100.86) .. (121.35,106.26) ;
%Shape: Boxed Bezier Curve [id:dp044320762840582395] 
\draw [color={rgb, 255:red, 37; green, 37; blue, 37 }  ,draw opacity=0.5 ][line width=1.5]    (106.7,116.61) .. controls (106.83,115.89) and (111.4,120.81) .. (107.87,124.74) .. controls (104.35,128.68) and (87.01,126.84) .. (89.65,117.39) .. controls (92.3,107.94) and (109.13,106.7) .. (115.81,112.67) .. controls (122.49,118.64) and (119.72,127.78) .. (116.21,130.6) .. controls (112.71,133.42) and (97.92,138.13) .. (82.42,132.72) ;

%Curve Lines [id:da9032308604670356] 
\draw [color={rgb, 255:red, 37; green, 37; blue, 37 }  ,draw opacity=0.5 ][line width=1.5]    (316.07,191.38) .. controls (315.94,192.09) and (311.37,187.18) .. (314.9,183.24) .. controls (318.43,179.3) and (335.77,181.14) .. (333.12,190.59) .. controls (330.48,200.04) and (313.65,201.28) .. (306.96,195.31) .. controls (300.28,189.34) and (303.06,180.21) .. (306.56,177.39) .. controls (310.07,174.56) and (324.85,169.86) .. (340.35,175.26) ;
%Shape: Boxed Bezier Curve [id:dp02146795795088674] 
\draw [color={rgb, 255:red, 37; green, 37; blue, 37 }  ,draw opacity=0.5 ][line width=1.5]    (325.7,185.61) .. controls (325.83,184.89) and (330.4,189.81) .. (326.87,193.74) .. controls (323.35,197.68) and (306.01,195.84) .. (308.65,186.39) .. controls (311.3,176.94) and (328.13,175.7) .. (334.81,181.67) .. controls (341.49,187.64) and (338.72,196.78) .. (335.21,199.6) .. controls (331.71,202.42) and (316.92,207.13) .. (301.42,201.72) ;

% Text Node
\draw (86.4,10.2) node [anchor=north west][inner sep=0.75pt]  [font=\normalsize] [align=left] {Input gauges};
% Text Node
\draw (303,84.2) node [anchor=north west][inner sep=0.75pt]  [font=\normalsize] [align=left] {Evaluation gauges};
% Text Node
\draw (298.4,266.8) node [anchor=north west][inner sep=0.75pt]  [font=\normalsize] [align=left] {Context gauges};
% Text Node
\draw (298.4,287) node [anchor=north west][inner sep=0.75pt]  [font=\normalsize] [align=left] {Holdout gauges};
% Text Node
\draw (9.6,308) node [anchor=north west][inner sep=0.75pt]  [font=\normalsize,rotate=-270] [align=left] {b) Densification};
% Text Node
\draw (9.6,144.8) node [anchor=north west][inner sep=0.75pt]  [font=\normalsize,rotate=-270] [align=left] {a) Hyperlocal};

\end{tikzpicture}